\documentclass[apj,twocolumn]{aastex63}
\usepackage{natbib}
\usepackage{textcomp}
\usepackage{graphicx}
\usepackage{amssymb}
\usepackage{rotating}
\usepackage{mathtools}
\usepackage{tablefootnote}
\usepackage{placeins}
\usepackage{comment}

\usepackage{lineno}

\usepackage{multirow}
\usepackage[toc,page]{appendix}
\usepackage{comment}
\usepackage[para,online,flushleft]{threeparttable}

\begin{document}

\title{Post-explosion evolution of core-collapse supernovae}

\author{M. Witt}
\affiliation{Institut f\"ur Kernphysik, Technische Universit\"at Darmstadt, 
Schlossgartenstr. 2, Darmstadt 64289, Germany}

\author{A. Psaltis}
\affiliation{Institut f\"ur Kernphysik, Technische Universit\"at Darmstadt, 
Schlossgartenstr. 2, Darmstadt 64289, Germany}

\author{H. Yasin}
\affiliation{Institut f\"ur Kernphysik, Technische Universit\"at Darmstadt, 
Schlossgartenstr. 2, Darmstadt 64289, Germany}

\author{C. Horn}
\affiliation{Institut f\"ur Kernphysik, Technische Universit\"at Darmstadt, 
Schlossgartenstr. 2, Darmstadt 64289, Germany}

\author{M. Reichert}
\affiliation{Institut f\"ur Kernphysik, Technische Universit\"at Darmstadt, 
Schlossgartenstr. 2, Darmstadt 64289, Germany}
\affiliation{Departament d'Astronomia i Astrof\'isica, Universitat de Val\`encia, Edifici d'Investigaci\'o Jeroni Munyoz, C/ Dr. Moliner, 50, E-46100 Burjassot (Val\'encia), Spain}

\author[0000-0001-5168-6792]{T. Kuroda}
\affiliation{Max-Planck-Institut f{\"u}r Gravitationsphysik, Am M{\"u}hlenberg 1, D-14476 Potsdam-Golm, Germany}

\author{M. Obergaulinger}
\affiliation{Departament d'Astronomia i Astrof\'isica, Universitat de Val\`encia, Edifici d'Investigaci\'o Jeroni Munyoz, C/ Dr. Moliner, 50, E-46100 Burjassot (Val\'encia), Spain}

\author{S. M. Couch}
\affiliation{Department of Physics and Astronomy, Michigan State University, East Lansing, MI 48824, USA}
\affiliation{Department of Computational Mathematics, Science, and Engineering, Michigan State University, East Lansing, MI 48824, USA}
\affiliation{National Superconducting Cyclotron Laboratory, Michigan State University, East Lansing, MI 48824, USA}
\affiliation{Joint Institute for Nuclear Astrophysics-Center for the Evolution of the Elements, Michigan State University, East Lansing, MI 48824, USA}

\author[0000-0002-6995-3032]{A. Arcones}
\affiliation{Institut f\"ur Kernphysik, Technische Universit\"at Darmstadt, 
Schlossgartenstr. 2, Darmstadt 64289, Germany}
\affiliation{GSI Helmholtzzentrum f\"ur Schwerionenforschung GmbH, Planckstr. 1, Darmstadt 64291, Germany}

\date{\today}

\begin{abstract}
We investigate the post-explosion phase in core-collapse supernovae with 2D hydrodynamical simulations and a simple neutrino treatment. The latter allows us to perform 46 simulations and follow the evolution of the 32 successful explosions during several seconds. We present a broad study based on three progenitors (11.2~$M_\odot$, 15~$M_\odot$, and 27~$M_\odot$), different neutrino-heating efficiencies, and various rotation rates. We show that the first seconds after shock revival determine the final explosion energy, remnant mass, and properties of ejected matter. Our results suggest that a continued mass accretion increases the explosion energy even at late times. We link the late-time mass accretion to initial conditions such as  rotation strength and shock deformation at explosion time. Only some of our simulations develop a neutrino-driven wind that survives for several seconds. This indicates that neutrino-driven winds are not a standard feature expected after every successful explosion. Even if our neutrino treatment is  simple, we estimate  the nucleosynthesis of the exploding models for the 15~$M_\odot$ progenitor after correcting the neutrino energies and luminosities to get a more realistic electron fraction.
\end{abstract}

\section{Introduction}
\label{sec:intro}

Core-collapse supernovae (CCSN) are exciting astrophysical events linked to a broad range of physics, expanding from the explosions themselves to neutrinos and nuclear physics and the high density equation of state, including also gravitational waves and nucleosynthesis. Depending on the aspect in focus, different kinds of hydrodynamical simulations are required. Ideally, one would perform 3D, fully general relativistic  magneto-hydrodynamic simulations with accurate neutrino transport for many progenitor stars, rotation rates, and magnetic field configurations and strengths, and follow the evolution during several seconds after the explosion. This is clearly computationally impossible today, therefore some aspects may be sacrificed to gain insights \citep[see e.g.,][]{Janka:2012, Kotake:2012, Mueller:2019_2, Burrows:2021}. Here, we focus on the long-time evolution of several seconds after the explosion and investigate the dynamics, accretion, explosion energy, and approximate nucleosynthesis and their dependency on neutrino heating, rotation, and progenitor. Therefore, we perform 2D Newtonian simulations with a simple neutrino treatment, namely gray leakage. This allows us to describe and understand important trends occurring during the first seconds after the explosions. However, our simplifications prevent us to make solid quantitative conclusions.

The long-time evolution has been considered in many nucleosynthesis studies of supernova yields. Originally spherically symmetric explosions were artificially triggered by pistons \citep{Woosley:1995} or thermal-energy bombs \citep{Thielemann:1996, Nagataki:1998,Nomoto:2006}. In the last decades, there have been numerous studies based on spherically symmetric simulations with enhanced neutrino energy deposition
\cite[e.g.,][]{Scheck:2006, Arcones:2007,OConnor:2010,Suwa:2011, Ugliano:2012, Perego:2015, Pejcha:2015,Ertl:2016,Sukhbold:2016,Mueller:2016_1, Curtis:2019,Ebinger:2019,Ebinger:2020,Couch:2020}. However, there are important multidimensional effects such as convection that impact the nucleosynthesis, explosion energy, explosion morphology \citep[see e.g.,][ and references therein]{Janka:2012,Kotake:2012, Mueller:2019_2, Burrows:2021}, and neutrino-driven wind \citep{Arcones:2011}. Recently, it has been possible to perform 2D simulations with good neutrino transport, to follow the evolution for a few seconds after the explosion and to study the supernova nucleosynthesis \cite[see e.g., ][]{Harris:2017, Eichler:2018, Wanajo:2018, Sieverding:2020,Reichert:2021}. However, these studies cover a reduced number of models and often only for short times after explosion.  Models with longer times usually require a mapping to a larger grid with simpler input physics after shock revival \citep{Wongwathanarat:2015,Mueller:2018,Stockinger:2020,Bollig:2020}

Rotation is an additional and important ingredient that affects not only the explosion but also the long-time evolution. The impact of rotation on the explosion has been extensively studied based on 2D \citep{LeBlanc:1970,Fryer:2000,Kotake:2003} \citep[see. e.g, ][and references therein]{Marek:2009, Suwa:2010,Blondin:2017, Summa:2018,Vartanyan:2018} and 3D simulations \citep{Kuroda:2014,Nakamura:2014, Moesta:2014,Takiwaki:2016,Blondin:2017,Summa:2018}. In this paper, we also show how rotation impacts  the evolution during the first seconds after the explosion, by affecting the shock morphology and the mass accretion onto the proto-neutron star (PNS).

This paper is organised as follows. In Sect.~\ref{sec:method}, we present our simulation setup, initial conditions, and nucleosythesis network. An overview of the models and their evolution during the first second after bounce is discussed in Sect.~\ref{sec:exp}. We investigate the long-time evolution in Sect.~\ref{sec:longtime} including the evolution of the explosion energy (Sect.~\ref{sec:Eexp}), the impact of rotation and shock deformation (Sect.~\ref{sec:long-time_dshock}), and the evolution of accretion and neutrino-driven wind (Sect.~\ref{sec:long-time_observables}). An estimate of the nucleosynthesis is given in Sect.~\ref{sec:nucleosynthesis}. We conclude in Sect.~\ref{sec:conclusions}.

\section{Simulations and nucleosynthesis}
\label{sec:method}

\subsection{Hydrodynamics and neutrinos}
\label{sec:hydro_neut}

We employ the multiphysics FLASH code \citep{Fryxell:2000,Dubey:2009} to perform 2D (cylindrical geometry), CCSN simulations. The domain size for all simulations is $3.2 \cdot 10^{10}$~cm along the cylindrical axis and $1.6 \cdot 10^{10}$~cm perpendicular to it. Using the adaptive mesh refinement in FLASH \citep{MacNeice:2000}, we achieve a maximum resolution of $\approx 488$~m. Our setup is similar to previous core-collapse supernova studies with FLASH, see e.g., \cite{Couch:2014,Couch:2015_1}. Self-gravity is calculated with a multipole approximation \citep{Couch:2013_2} of Poisson's equation, without the effective general relativistic potential that was used in \citet{OConnor:2018_1}.

Before bounce, the deleptonization scheme of \cite{Liebendoerfer:2005_1} is used. After bounce, neutrinos are described by a gray, three-flavor neutrino leakage scheme with a ray-by-ray approximation, as in \cite{OConnor:2010, Couch:2014}. In order to facilitate  the explosions,  the neutrino heating  is artificially enhanced by a factor, $f_\mathrm{heat}$, that was introduced in previous studies \citep{OConnor:2010, Couch:2014, Couch:2015_1}. Here, we adjust slightly the implementation  to have the enhanced heating only in the gain layer and until the shock has reached a radius of $1000$~km.

Leakage schemes are not as accurate as the more sophisticated neutrino transport methods  \cite[see e.g., ][]{Just:2015, OConnor:2015, Kuroda:2016, Pan:2019,Glas:2019_1}. However, broad studies of multi-dimensional, long-time simulations up to $10$~s after bounce are yet not feasible with state-of-the-art transport methods, without sacrificing resolution. A consequence of the more approximate leakage scheme is an incorrect electron fraction, typically leaning towards more neutron-rich values \citep{Dessart:2009, Pan:2019}. We  discuss corrections to the electron fraction for nucleosynthesis calculations in Sec.~\ref{sec:nucleosynthesis}.

We use the LS220 equation of state (EOS) of \cite{Lattimer:1991} for the high density regions. In the outer layers of the progenitor and at late times, the density reaches very low values ($\rho \lesssim 10~\mathrm{g}\,\mathrm{cm}^{-3}$), therefore, we use a hybrid EOS approach with a transition to the Helmholtz EOS \citep{Timmes:1999, Timmes:2000, Fryxell:2000} at around  $5-1 \cdot 10^{5}~\mathrm{g}~\mathrm{cm}^{-3}$.

\subsection{Progenitor models and rotation}
\label{sec:pog_rot}

In this study, we use three different progenitors with ZAMS mass 11.2~$M_\odot$, 15~$M_\odot$, and 27~$M_\odot$  namely the s11.2, s15.0 and s27.0 models from \cite{Woosley:2002}. All three progenitor models are non-rotating. As we aim to explore the effect of rotation  on the explosion phase and on the long-time evolution, we superimpose a parametric rotational profile of the form
\begin{equation}
 \label{eq:rotation_parametric}
 \Omega(r) \, = \, \Omega_0 \cdot \frac{1}{1+(r/r_A)^2} \, .
\end{equation}
We set the characteristic radius $r_A$ to a fixed value of $3000$~km \citep[approximately the extent of the Fe and Si core in the s15 progenitor, see][]{Mueller:2004,Buras:2006} and use $\Omega_0$ as a free parameter to adjust the rotation strength. In addition to a non-rotating case ($\Omega_0 = 0$), we use six different rotation strengths:  $\Omega_0$~=~(0.01, 0.03, 0.06, 0.10, 0.20, 0.30)~$\cdot~(2\pi~\mathrm{rad \,s}^{-1})$. The three first values correspond to moderately rotating models \citep[compare to, e.g.,][]{Heger:2005,Ott:2006,Burrows:2007_2} and the three last ones to rapidly rotating.

We label our models with their corresponding progenitor, heating factor, and rotation strength. For example, the label s15\_F130\_R006 refers to a simulation of the s15.0 progenitor, with $f_\mathrm{heat} = 1.3$, and $\Omega_0 = 0.06 \cdot 2\pi$~rad/s.

\subsection{Tracer particles and nucleosynthesis network}
\label{sec:tracer_method}

In order to estimate the nucleosynthesis, we need the Lagrangian evolution of the matter ejected. Since the FLASH code is Eulerian, we use a tracer particle scheme  to track individual fluid elements (also called tracer particles, trajectories, mass elements). The tracers are not included in the simulation but calculated from the output (see e.g., \citet{Harris:2017} for detailed discussion). All tracer particles have equal mass and are distributed proportional to the density \cite[see][for tracers in FLASH code]{Dubey:2012}. For  the s15.0 progenitor, we initialize 21,281 particles at the beginning of the simulation. This corresponds to a mass of $1.76~\cdot~10^{-4}~M_\odot$ per tracer that is similar to the ``medium resolution'' case in \citet{Nishimura:2015}.  The amount of matter ejected by the tracers agrees within 1\% with the unbound material obtained directly from the hydrodynamics.

In addition to the evolution of density and temperature along the tracers, we also need the electron fraction. However, this is poorly determined in our simulations because we use a gray neutrino leakage scheme. Therefore, we correct the electron fraction to  estimate the nucleosynthesis within some uncertainties (see Sect.~\ref{sec:nucleosynthesis} for more details).

For the nucleosynthesis calculations, we use the nuclear reaction network code WinNet \citep{Winteler:2011,Winteler:2012}. The reaction rates include the JINA Reaclib compilation \citep{Cyburt:2010}, theoretical weak reactions from \citet{Langanke:2001}, and neutrino reactions from \citep[see also \citealt{Froehlich:2006b} for details about the neutrino reactions]{Langanke:2001b}. For all nucleosynthesis calculations, we evolve the electron fraction in nuclear statistical equilibrium (NSE) at a temperature of 20~GK and assume NSE down to 6.5~GK.

\section{Overview of models and explosions}
\label{sec:exp}
We present here all our models including 32 successful explosions (see Table~\ref{tab:model_overview} for an overview). We also include unsuccessful explosions in the table for completeness. These models were continued for 1-2~s after bounce but did not show signs of shock revival until then. First, we focus shortly on the explosion phase and discuss the long-time evolution in more detail later  (Sect.~\ref{sec:longtime}). We are aware that our results are only qualitative because the neutrino transport is approximated and we artificially increase the energy deposited by neutrinos to trigger explosions. Moreover, our simulations are 2D and this favours some artificial features and may amplify some of our conclusions, like the persistence of downflows. However, we are able to explore the long-time evolution for a big sample of models and we find trends and correlations that are robust and are present also in 3D simulations with better neutrino treatment \citep[e.g., see][]{Mueller:2015_2,Janka:2016,Summa:2018,Burrows:2019,Mueller:2019_1,Vartanyan:2019_1,Stockinger:2020,Bollig:2020}. 
 
\begin{table*}
\caption{Overview of models}
\label{tab:model_overview}
\centering
\begin{threeparttable}
\begin{tabular}{cccccccccccc}
    name & prog. & $f_\mathrm{heat}$ & $\Omega_0$  & $t_\mathrm{end}$ & $t_\mathrm{exp}$\tnote{a} & $d_\mathrm{shock}$\tnote{b} & $E_\mathrm{exp}^\mathrm{100ms}$\tnote{  c} & $E_\mathrm{exp}^\mathrm{final}$ & $M_\mathrm{ej}$\tnote{d} & $M_\mathrm{PNS}$ & $t_\mathrm{wind}$\tnote{e} \\
    &&&[2$\pi$ rad/s]&[s] &[s] & & [B] & [B] & [$M_\odot$] & [$M_\odot$] & [s] \\ 
    \hline
    \hline
s15\_F100\_R000 & s15.0 & 1.00 & 0.00 & 1.55 & ... & ... & ... & ... & ... & 1.98 & ... \\ 
s15\_F105\_R000 & s15.0 & 1.05 & 0.00 & 1.54 & ... & ... & ... & ... & ... & 1.98 & ... \\ 
s15\_F105\_R003 & s15.0 & 1.05 & 0.03 & 1.55 & ... & ... & ... & ... & ... & 1.98 & ... \\ 
s15\_F105\_R020 & s15.0 & 1.05 & 0.20 & 1.45 & ... & ... & ... & ... & ... & 1.96 & ... \\ 
s15\_F108\_R000 & s15.0 & 1.08 & 0.00 & 8.96 & 1.07 & 1.26 & 0.04 & 1.16 & 0.76 & 2.16 & 0.00 \\ 
s15\_F110\_R000 & s15.0 & 1.10 & 0.00 & 8.14 & 0.93 & 1.06 & 0.04 & 1.41 & 0.90 & 2.10 & 0.00 \\ 
s15\_F110\_R001 & s15.0 & 1.10 & 0.01 & 1.63 & ... & ... & ... & ... & ... & 1.99 & ... \\ 
s15\_F110\_R003 & s15.0 & 1.10 & 0.03 & 1.54 & ... & ... & ... & ... & ... & 1.98 & ... \\ 
s15\_F110\_R020 & s15.0 & 1.10 & 0.20 & 1.46 & ... & ... & ... & ... & ... & 1.96 & ... \\ 
s15\_F115\_R000 & s15.0 & 1.15 & 0.00 & 9.25 & 0.92 & 1.01 & 0.04 & 1.34 & 0.93 & 2.20 & 0.00 \\ 
s15\_F115\_R003 & s15.0 & 1.15 & 0.03 & 1.61 & ... & ... & ... & ... & ... & 1.99 & ... \\ 
s15\_F120\_R000 & s15.0 & 1.20 & 0.00 & 7.89 & 0.28 & 0.94 & 0.14 & 1.21 & 0.86 & 2.03 & 0.00 \\ 
s15\_F120\_R001 & s15.0 & 1.20 & 0.01 & 6.74 & 0.30 & 0.87 & 0.19 & 1.20 & 0.76 & 1.95 & 0.00 \\ 
s15\_F120\_R003 & s15.0 & 1.20 & 0.03 & 9.31 & 0.36 & 0.29 & 0.15 & 0.70 & 1.28 & 1.80 & 2.44 \\ 
s15\_F120\_R006 & s15.0 & 1.20 & 0.06 & 6.03 & 0.38 & 0.77 & 0.11 & 0.63 & 0.88 & 1.86 & 0.00 \\ 
s15\_F120\_R010 & s15.0 & 1.20 & 0.10 & 1.61 & ... & ... & ... & ... & ... & 1.99 & ... \\ 
s15\_F120\_R020 & s15.0 & 1.20 & 0.20 & 1.46 & ... & ... & ... & ... & ... & 1.96 & ... \\ 
s15\_F130\_R000 & s15.0 & 1.30 & 0.00 & 9.26 & 0.23 & 0.59 & 0.15 & 0.79 & 1.42 & 1.75 & 1.73 \\ 
s15\_F130\_R001 & s15.0 & 1.30 & 0.01 & 4.60 & 0.22 & 0.37 & 0.27 & 1.00 & 0.82 & 1.85 & 0.00 \\ 
s15\_F130\_R003 & s15.0 & 1.30 & 0.03 & 8.63 & 0.23 & 0.37 & 0.22 & 0.75 & 1.52 & 1.68 & 3.48 \\ 
s15\_F130\_R006 & s15.0 & 1.30 & 0.06 & 4.17 & 0.24 & 0.25 & 0.28 & 0.73 & 0.59 & 1.87 & 0.00 \\ 
s15\_F130\_R010 & s15.0 & 1.30 & 0.10 & 9.08 & 0.28 & -0.00 & 0.24 & 0.64 & 1.40 & 1.72 & 5.10 \\ 
s15\_F130\_R020 & s15.0 & 1.30 & 0.20 & 9.25 & 0.32 & -0.13 & 0.19 & 0.77 & 1.34 & 1.75 & 2.46 \\ 
s15\_F130\_R030 & s15.0 & 1.30 & 0.30 & 9.30 & 0.36 & 0.20 & 0.15 & 0.67 & 1.26 & 1.76 & 3.60 \\ 
s15\_F150\_R000 & s15.0 & 1.50 & 0.00 & 7.96 & 0.17 & 0.64 & 0.30 & 1.21 & 1.13 & 1.87 & 0.00 \\ 
s15\_F150\_R001 & s15.0 & 1.50 & 0.01 & 7.69 & 0.17 & 0.68 & 0.36 & 1.12 & 1.31 & 1.71 & 0.16 \\ 
s15\_F150\_R003 & s15.0 & 1.50 & 0.03 & 9.43 & 0.21 & 0.24 & 0.40 & 1.09 & 1.59 & 1.69 & 3.37 \\ 
s15\_F150\_R006 & s15.0 & 1.50 & 0.06 & 9.03 & 0.17 & 0.19 & 0.36 & 0.91 & 1.56 & 1.69 & 5.34 \\ 
s15\_F150\_R010 & s15.0 & 1.50 & 0.10 & 8.00 & 0.21 & 0.22 & 0.31 & 0.73 & 1.55 & 1.69 & 4.81 \\ 
s15\_F150\_R020 & s15.0 & 1.50 & 0.20 & 8.94 & 0.22 & -0.21 & 0.37 & 0.83 & 1.57 & 1.67 & 4.81 \\ 
s15\_F150\_R030 & s15.0 & 1.50 & 0.30 & 9.40 & 0.21 & 0.18 & 0.16 & 0.81 & 1.42 & 1.69 & 3.80 \\ 
\hline
s11\_F100\_R000 & s11.2 & 1.00 & 0.00 & 4.36 & 0.16 & 0.36 & 0.04 & 0.36 & 0.41 & 1.33 & 0.00 \\ 
s11\_F100\_R003 & s11.2 & 1.00 & 0.03 & 4.43 & 0.25 & 0.19 & 0.04 & 0.44 & 0.33 & 1.39 & 0.00 \\ 
s11\_F100\_R020 & s11.2 & 1.00 & 0.20 & 4.63 & 0.25 & 0.61 & 0.02 & 0.64 & 0.31 & 1.40 & 0.00 \\ 
\hline
s27\_F105\_R000 & s27.0 & 1.05 & 0.00 & 3.79 & 1.42 & 0.93 & 0.05 & 0.67 & 0.27 & 2.22 & 0.00 \\ 
s27\_F105\_R003 & s27.0 & 1.05 & 0.03 & 1.82 & ... & ... & ... & ... & ... & 2.07 & ... \\ 
s27\_F105\_R020 & s27.0 & 1.05 & 0.20 & 1.73 & ... & ... & ... & ... & ... & 2.05 & ... \\ 
s27\_F110\_R000 & s27.0 & 1.10 & 0.00 & 3.06 & 0.31 & 0.63 & 0.07 & 0.33 & 0.17 & 1.93 & 0.00 \\ 
s27\_F110\_R003 & s27.0 & 1.10 & 0.03 & 1.30 & 0.54 & 0.39 & 0.05 & 0.23 & 0.09 & 1.87 & 0.00 \\ 
s27\_F110\_R020 & s27.0 & 1.10 & 0.20 & 2.06 & ... & ... & ... & ... & ... & 2.12 & ... \\ 
s27\_F120\_R000 & s27.0 & 1.20 & 0.00 & 2.89 & 0.29 & 0.72 & 0.07 & 0.65 & 0.39 & 2.02 & 0.00 \\ 
s27\_F120\_R003 & s27.0 & 1.20 & 0.03 & 1.79 & 0.34 & 0.59 & 0.10 & 0.55 & 0.19 & 1.91 & 0.00 \\ 
s27\_F120\_R020 & s27.0 & 1.20 & 0.20 & 2.11 & ... & ... & ... & ... & ... & 2.13 & ... \\ 
s27\_F130\_R000 & s27.0 & 1.30 & 0.00 & 3.47 & 0.21 & 0.73 & 0.07 & 1.24 & 0.55 & 2.09 & 0.00 \\ 
s27\_F130\_R003 & s27.0 & 1.30 & 0.03 & 2.85 & 0.20 & 0.55 & 0.10 & 0.90 & 0.38 & 2.00 & 0.00 \\ 
s27\_F130\_R020 & s27.0 & 1.30 & 0.20 & 3.82 & 0.29 & 0.16 & 0.07 & 0.46 & 0.23 & 1.92 & 0.00 \\ 
\hline
\end{tabular}
 \begin{tablenotes}
  \item[a] Post-bounce time when the maximum shock radius exceeds $600$~km
  \item[b] Explosion energy at $t_\mathrm{exp} + 100$~ms
  \item[c] Shock deformation parameter (Eq.~\ref{eq:dshock}) at $t_\mathrm{exp}$
  \item[d] Mass of ejected (unbound) matter at the end of the simulation
  \item[e] Total duration of neutrino-driven wind phases ($\dot{M}_\mathrm{acc,500km} = 0$)
 \end{tablenotes}
\end{threeparttable}
\end{table*}

\begin{figure}[ht]
 \centering
 \includegraphics[width=0.5\linewidth]{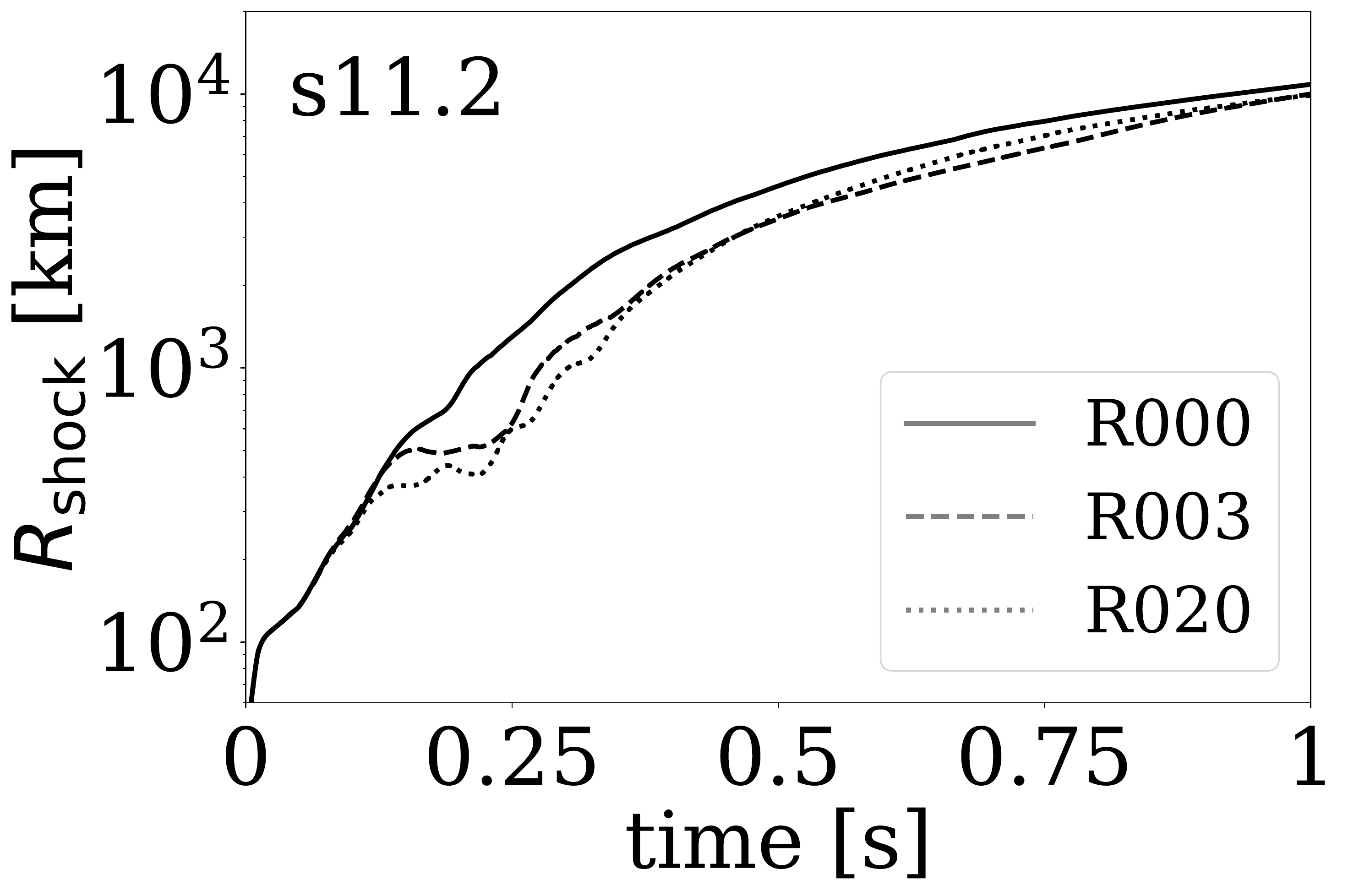}%
 \includegraphics[width=0.5\linewidth]{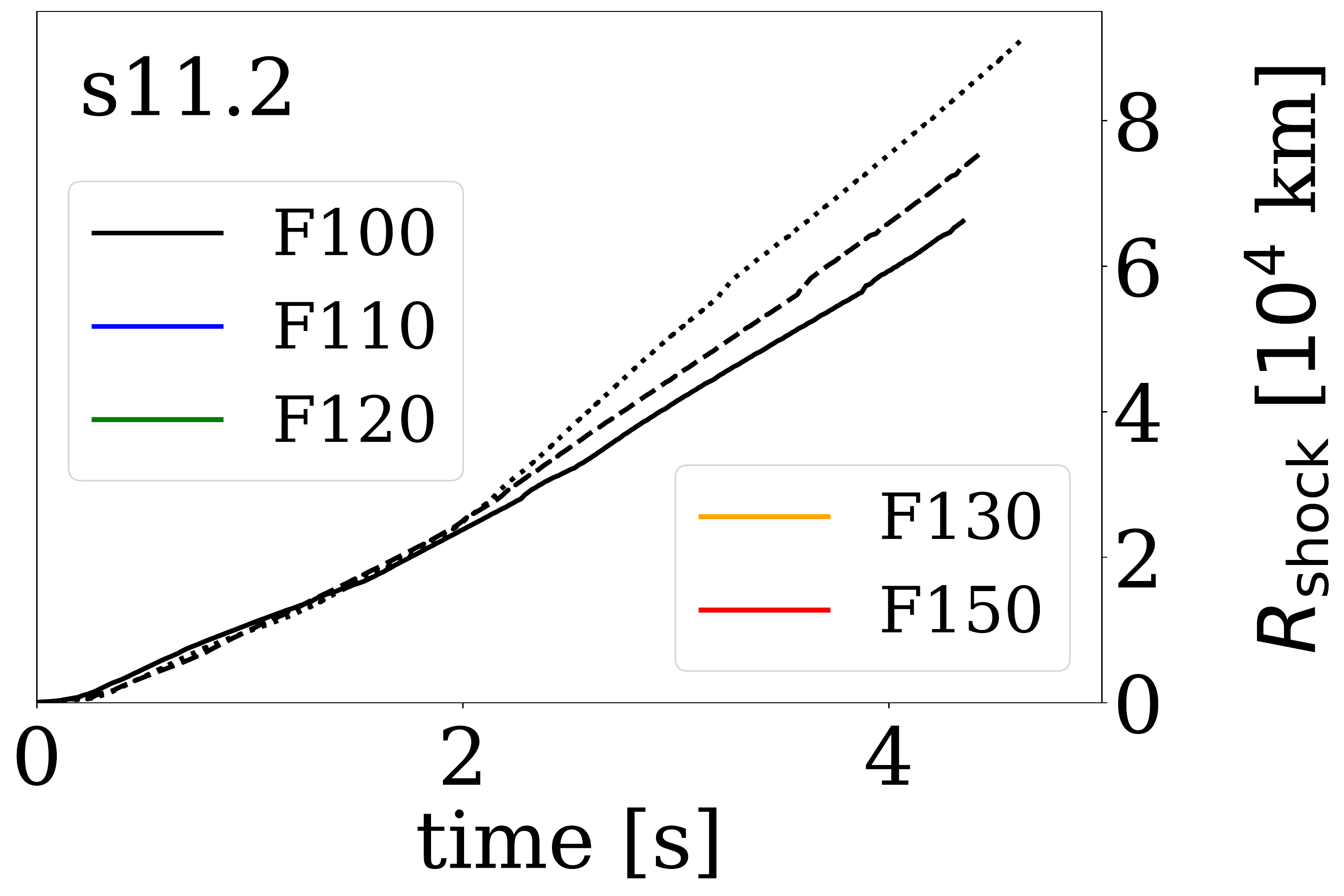}\\
  \includegraphics[width=0.5\linewidth]{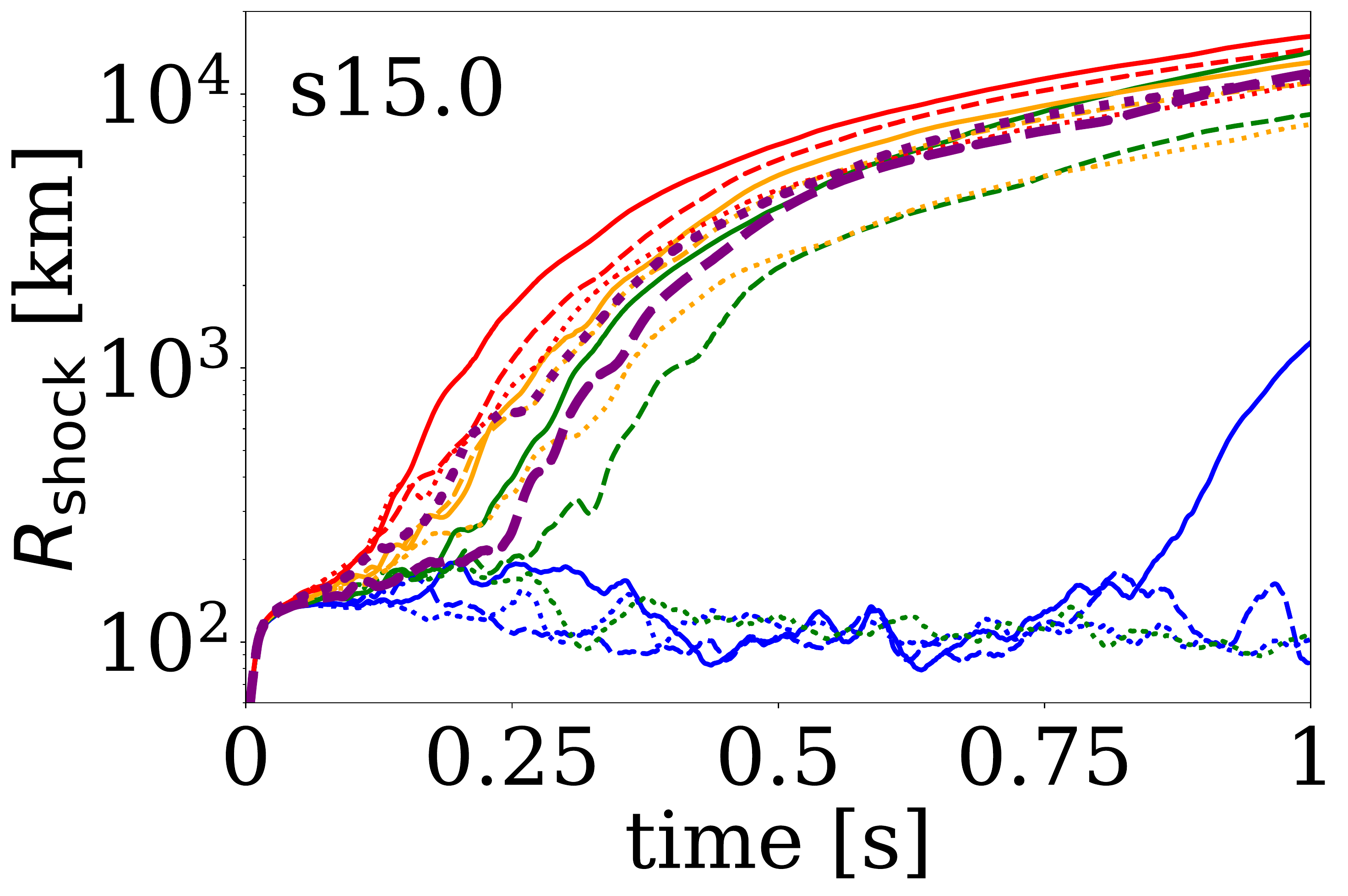}%
 \includegraphics[width=0.5\linewidth]{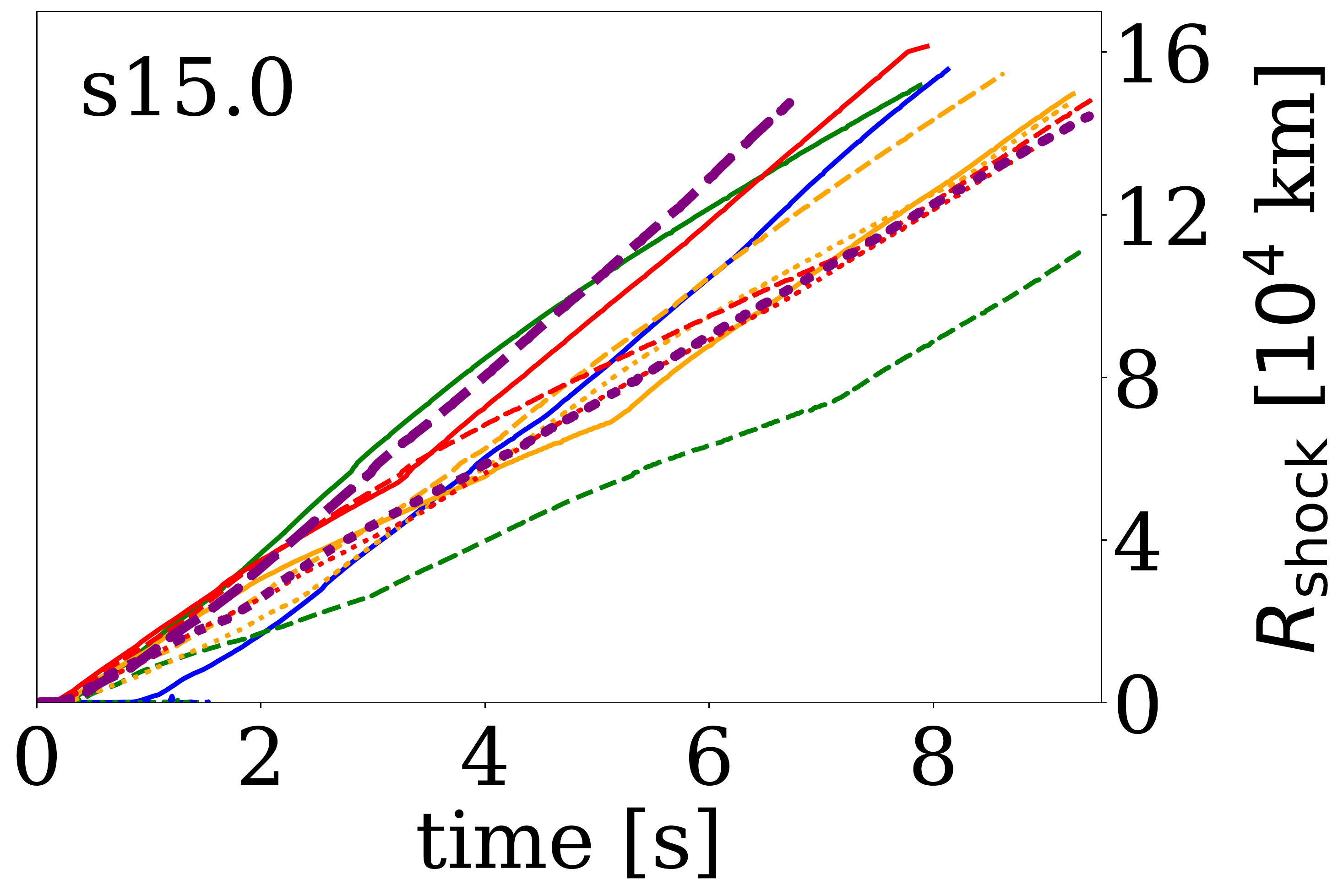}\\
  \includegraphics[width=0.5\linewidth]{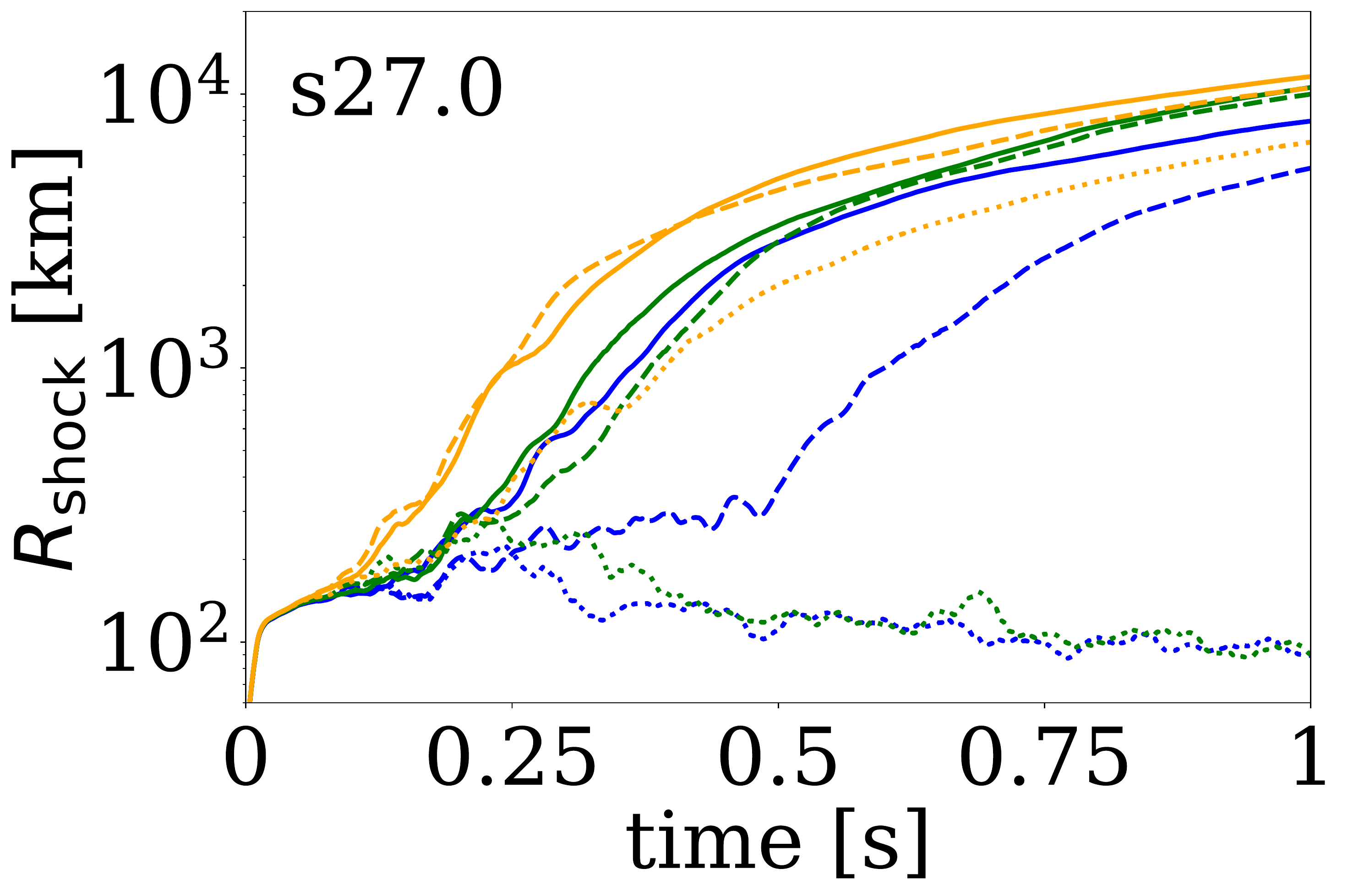}%
 \includegraphics[width=0.5\linewidth]{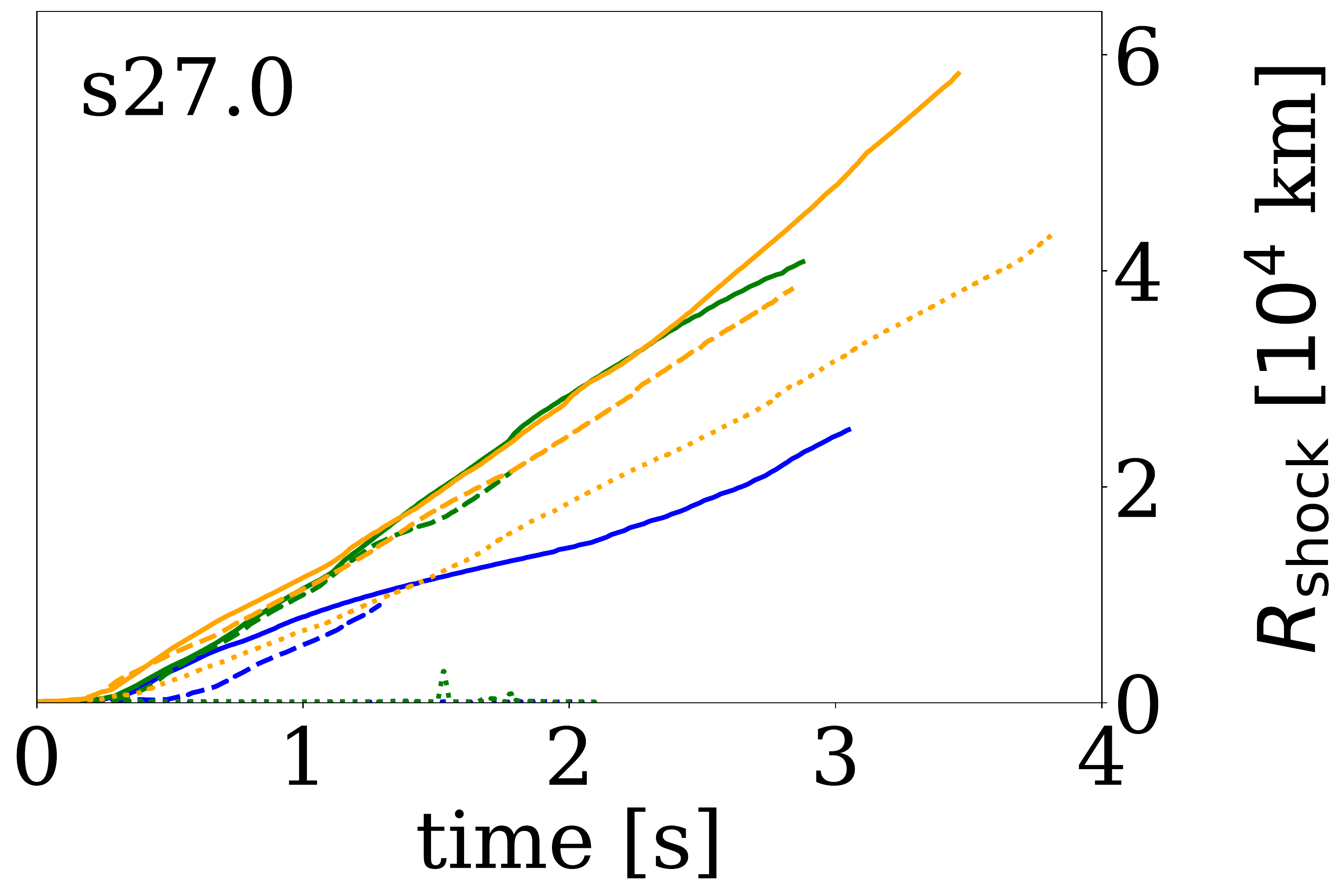}
 \caption{Early (left column) and long-time (right column) evolution of the maximum shock radii for  the three progenitors  (s11.2, s15.0, and s27.0 in top, middle, and bottom panels, respectively). Different rotation rates are indicated by different line style  an various heating factor by colours, as given in the figures. The dashed and dotted purple lines represent the models s15\_F120\_R001 and s15\_F150\_R030, respectively.}
 \label{fig:rshk}
\end{figure}

The evolution of the shock radius is shown in Fig.~\ref{fig:rshk} for the three progenitors and for three rotation rates. The lightest progenitor, s11.2, explodes without additional neutrino heating within $0.25$~s for all variations of the rotation. Therefore, we only use $f_\mathrm{heat} = 1$ for this progenitor. For the heavier progenitors, the threshold for shock revival in non-rotating models is $f_\mathrm{heat} = 1.08$ and $1.05$ for s15.0 and s27.0, respectively. This is in agreement with estimates from progenitor compactness \citep{OConnor:2011}, which has values of $\xi_\mathrm{1.75} = 0.07$, $0.54$ and $0.53$ for the s11.2, s15.0 and s27.0 models, respectively \citep{Pan:2016,Summa:2016}. The explosion time varies depending on the progenitor and the heating factor and it is also affected by rotation.

\begin{figure*}[ht]
 \centering
 \includegraphics[width=0.9\linewidth]{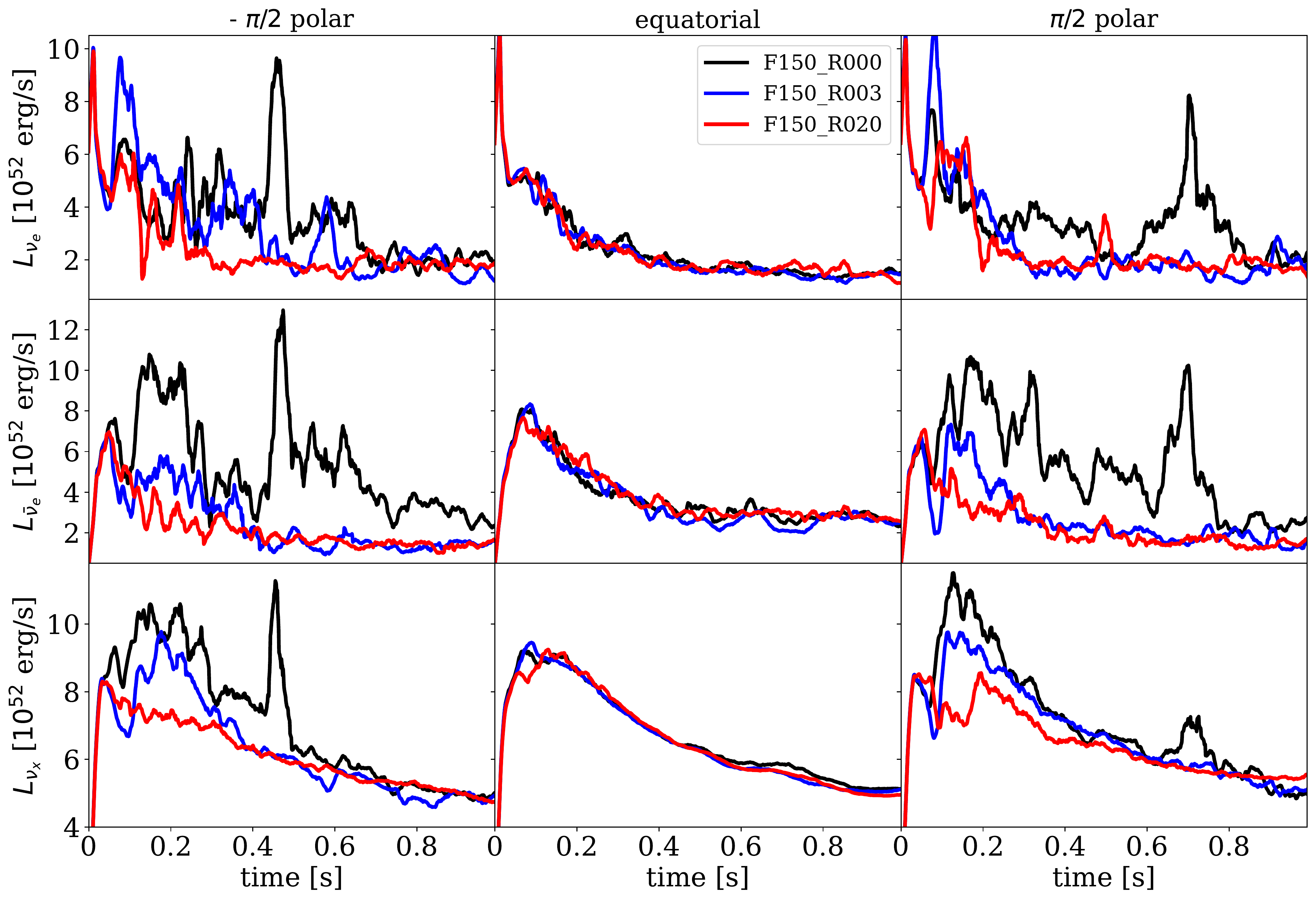}
 \caption{Neutrino luminosities for non-rotating (black), moderately (blue), and rapidly rotating (red) for the s15.0 progenitor with a heating factor $f_\mathrm{heat} = 1.5$. The columns correspond to observer directions along the poles (left and right panels) and the equatorial plane (center panels). The different rows show the luminosity for  electron neutrinos (top), antineutrinos (middle), and heavy-lepton neutrinos (bottom).}
 \label{fig:rotation_angular_lumAll}
\end{figure*}

In the presence of rotation, we notice that shock revival generally occurs later, or fails overall \citep[see also, e.g., ][]{Fryer:2000,Thompson:2004,Summa:2018,Obergaulinger:2020}. This can be understood from a reduction of mass accretion through the stalled shock because of centrifugal forces. This leads to a diminished accretion luminosity and ultimately takes away support for the stalled shock.  Moreover, rotation contributes to shift matter from the poles towards the equatorial plane and this has an impact on the  neutrino luminosities in different directions. Figure~\ref{fig:rotation_angular_lumAll} indicates that, in the direction of the poles, the luminosities for rotating models are considerably smaller, in accordance with the lower density and mass accretion in this region. Notice that in 3D simulations and with magnetic fields, the smaller mass accretion rates and luminosities are often found in the equatorial plane, but the integrated luminosities are also smaller in the context of rapid rotation \citep{Summa:2018,Obergaulinger:2020}. There is no significant change in the equatorial plane. Both effects lead to a smaller integrated neutrino heating rate for rotating models. Anisotropic emission of neutrinos in rotating CCSN has already been reported in early multi-dimensional simulations \citep{Fryer:2000,Kotake:2003,Buras:2003,Thompson:2004,Marek:2009}. The described behaviour may be artificially enhanced in two-dimensional simulations, although, modern simulations in three dimensions also predict smaller luminosities in rapidly rotating cases \citep{Summa:2018}. However, non-axisymmetric spiral modes of the standing accretion shock instability (SASI) can increase the mass in the gain region and compensate the lack of specific heating, leading to a fundamentally different explosion behavior in 3D \citep{Nakamura:2014,Summa:2018,Kuroda:2020}.

\section{Long-time evolution}
\label{sec:longtime}
The 32 exploding models are evolved for several seconds (see Table~\ref{tab:model_overview}) and we define the ``long-time'' phase starting one second after the explosion, i.e., $t > t_\mathrm{exp} + 1$~s. The long-time evolution is characterized by a continuous increase of the explosion energy due to long-lasting accretion onto the proto-neutron star. This accretion depends on the formation and evolution of downflows and it is correlated with rotation and to the morphology of the early explosion phase. In the following, we discuss the generation of explosion energy (Sect.~\ref{sec:Eexp}), how this is affected by rotation and by the explosion morphology (Sect.~\ref{sec:long-time_dshock}), and the evolution of key quantities during the long-time evolution phase (Sect.~\ref{sec:long-time_observables}).

\subsection{Explosion-energy generation}
\label{sec:Eexp}

\begin{figure}[ht]
 \centering
 \includegraphics[width=0.5\linewidth]{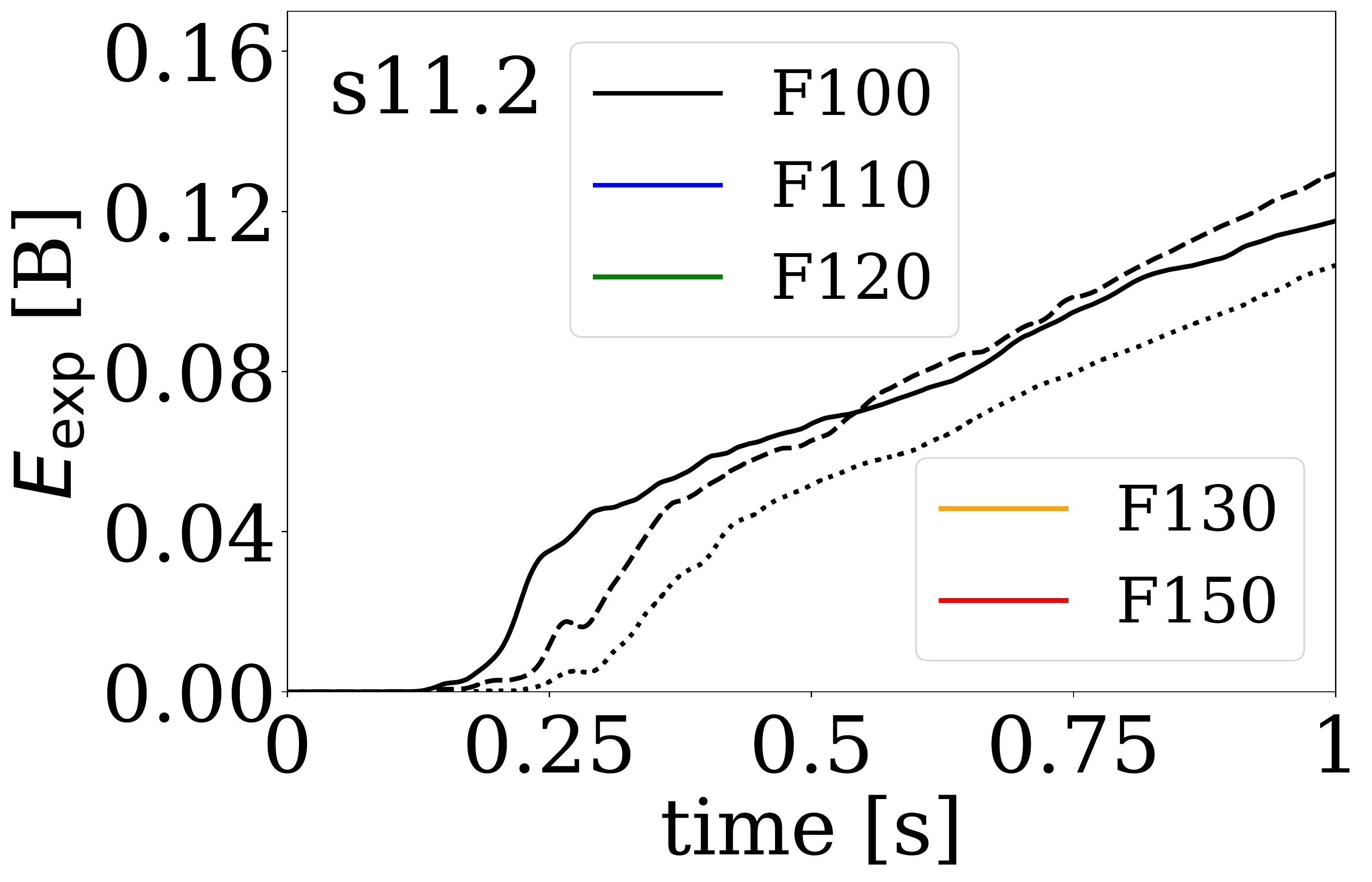}%
 \includegraphics[width=0.5\linewidth]{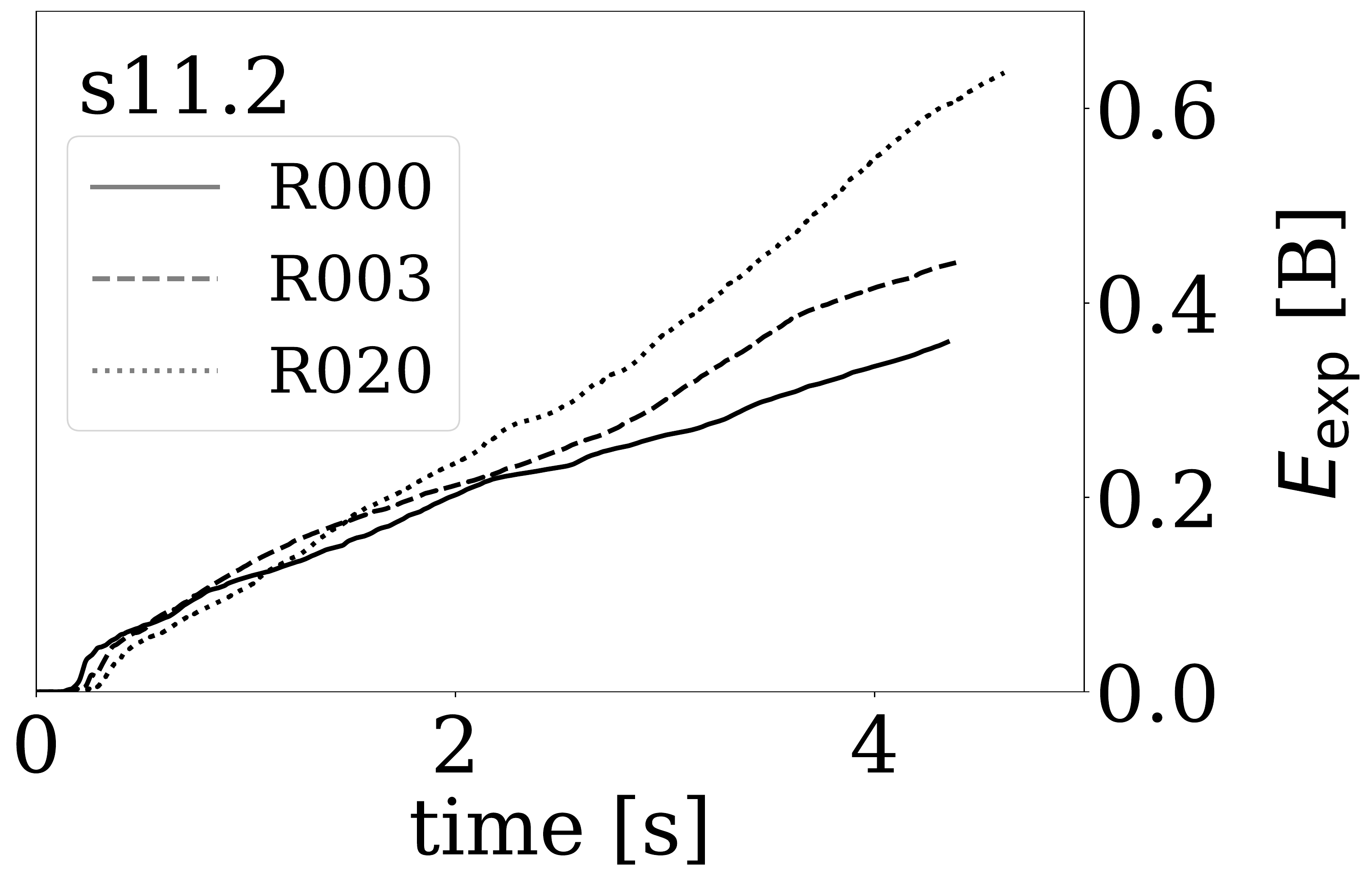}\\
  \includegraphics[width=0.5\linewidth]{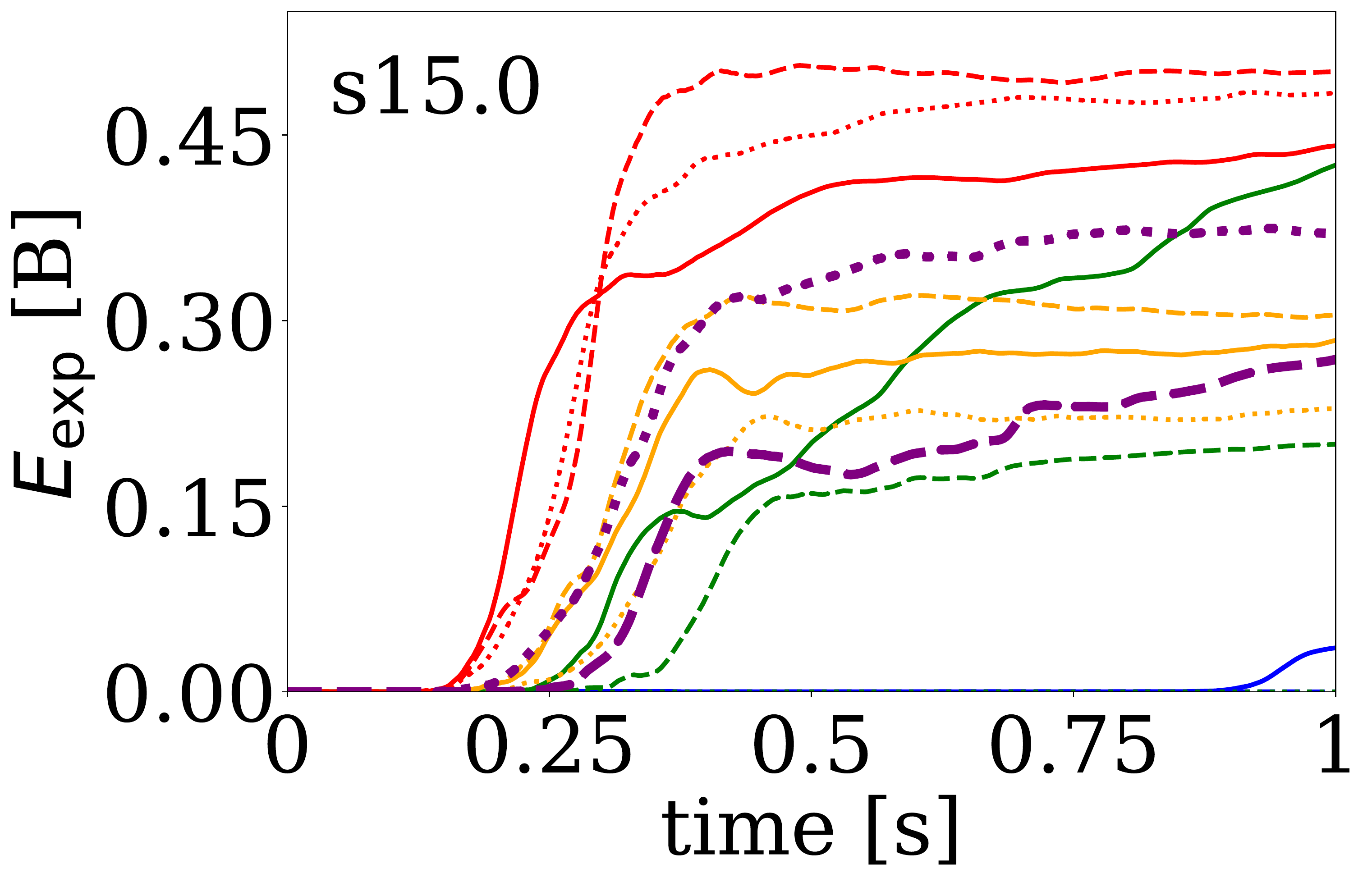}%
 \includegraphics[width=0.5\linewidth]{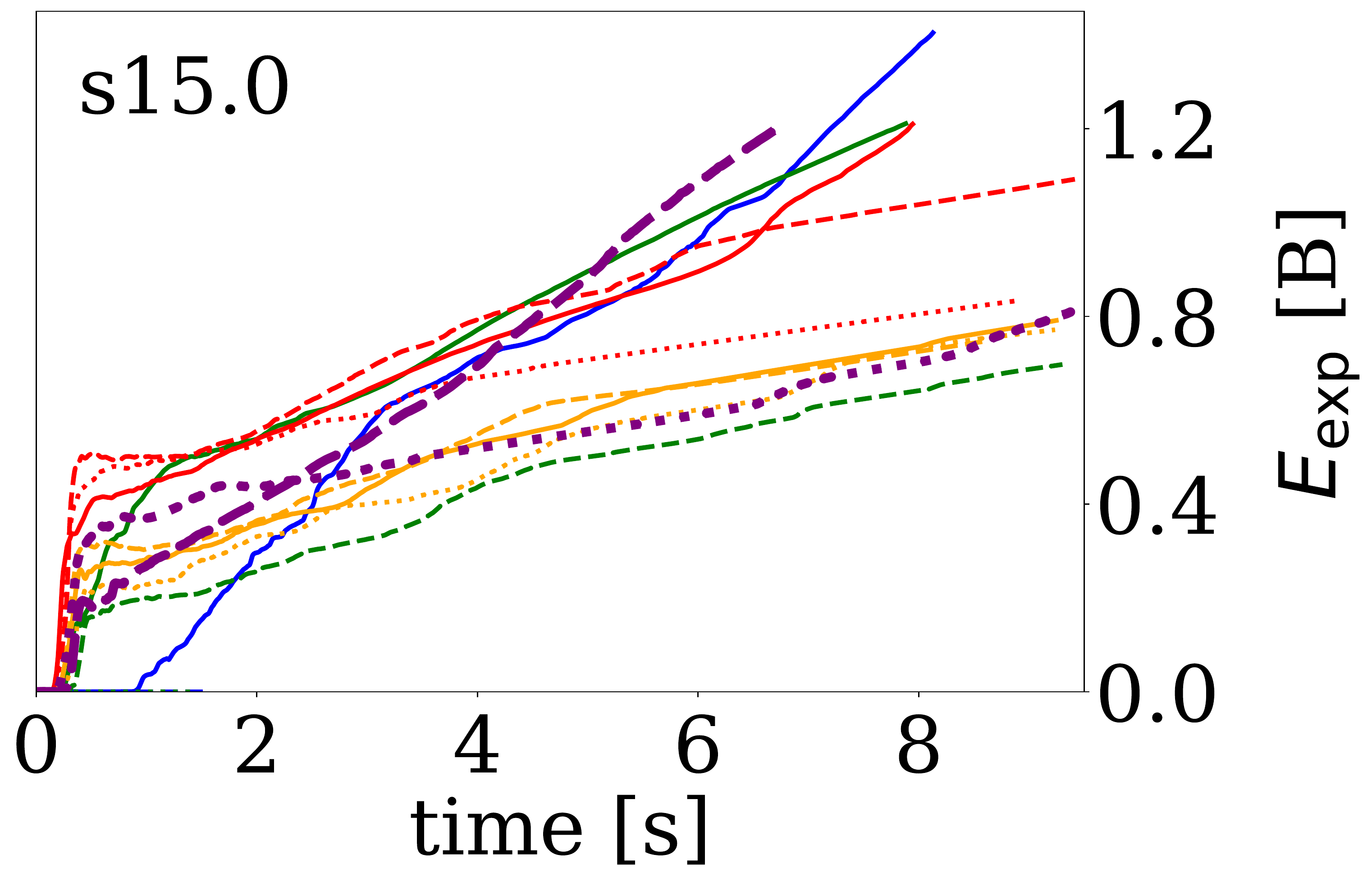}\\
  \includegraphics[width=0.5\linewidth]{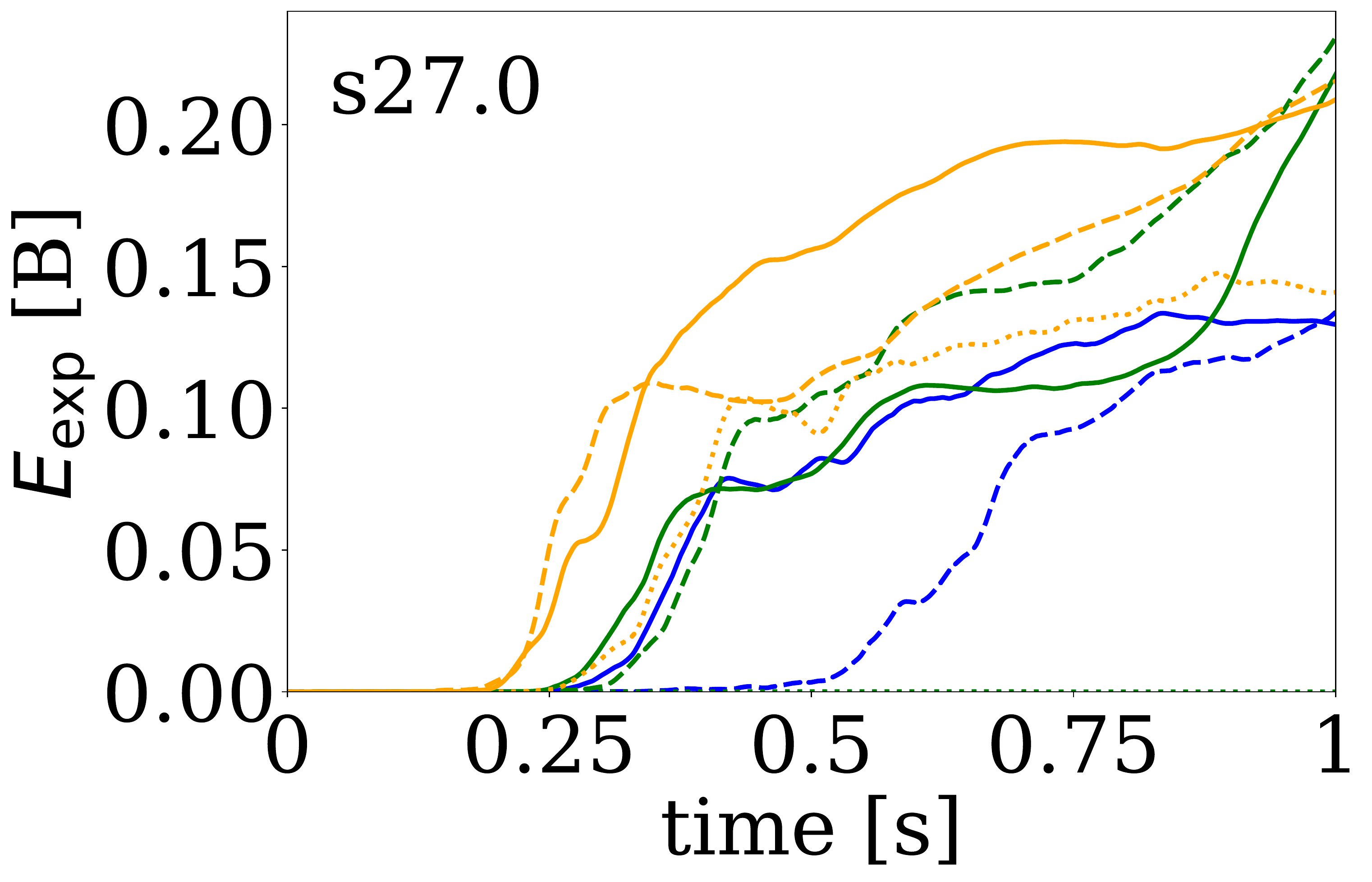}%
 \includegraphics[width=0.5\linewidth]{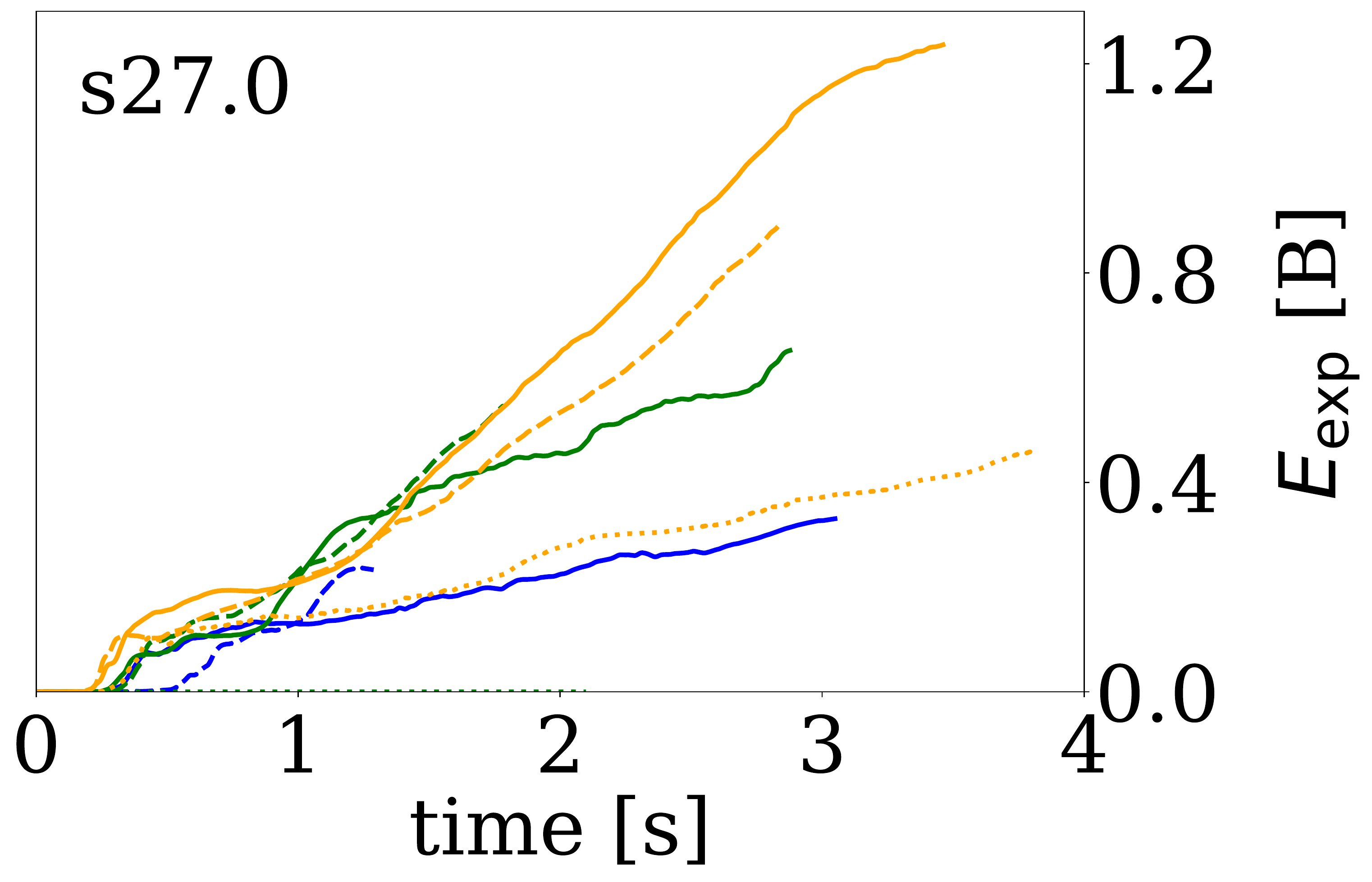}
 \caption{Same as in Fig.~\ref{fig:rshk}, but for the explosion energy. The dashed and dotted purple lines represent the models s15\_F120\_R001 and s15\_F150\_R030, respectively.}
 \label{fig:eexp}
\end{figure}

The explosion energy ($E_\mathrm{exp}$) is calculated by adding the energy of all unbound matter\footnote{In the calculation of the explosion energy we do not consider the contribution of the outer layers  \cite[see e.g.,][]{Bruenn:2013,Bruenn:2016}. We estimate that this ``overburden'' contribution will be around $E_\mathrm{ov} \ll 0.1$~B in our models, which is not important for our overall qualitative description.}  and it is shown in Fig.~\ref{fig:eexp} for the three progenitors including different heating factors and rotation rates. We note that our definition corresponds to the ``diagnostic'' explosion energy, in the sense that it is not the final, saturated value. Shortly after the explosion, $E_\mathrm{exp}$ rapidly grows and in some cases it looks like it saturates and stays constant. However, for all models the explosion energy continues slowly increasing during seconds due to the long-lasting accretion \citep[see also][]{Bruenn:2013,Bruenn:2016, Nakamura:2015, Nakamura:2016, Mueller:2017, Summa:2016, Summa:2018, Harris:2017, OConnor:2018_1, Vartanyan:2018, Vartanyan:2019_1, Glas:2019_1, Mueller:2019_1, Stockinger:2020, Bollig:2020}. This is in contrast to one-dimensional models, that by definition cannot account for downflows and the explosion energy saturates promptly and stays constant \citep[e.g., see][]{Arcones:2007,Perego:2015,Couch:2020}. Even if long-lasting downflows in 2D simulations may become stable and artificially stay for too long time, there are also 3D simulations showing downflows and accretion at late times \citep{Burrows:2019,Vartanyan:2019_1,Stockinger:2020,Bollig:2020}. With our large set of models, we can investigate the  impact of different key aspects on the explosion energy.

\begin{figure}[ht]
 \centering
 \includegraphics[width=1\linewidth]{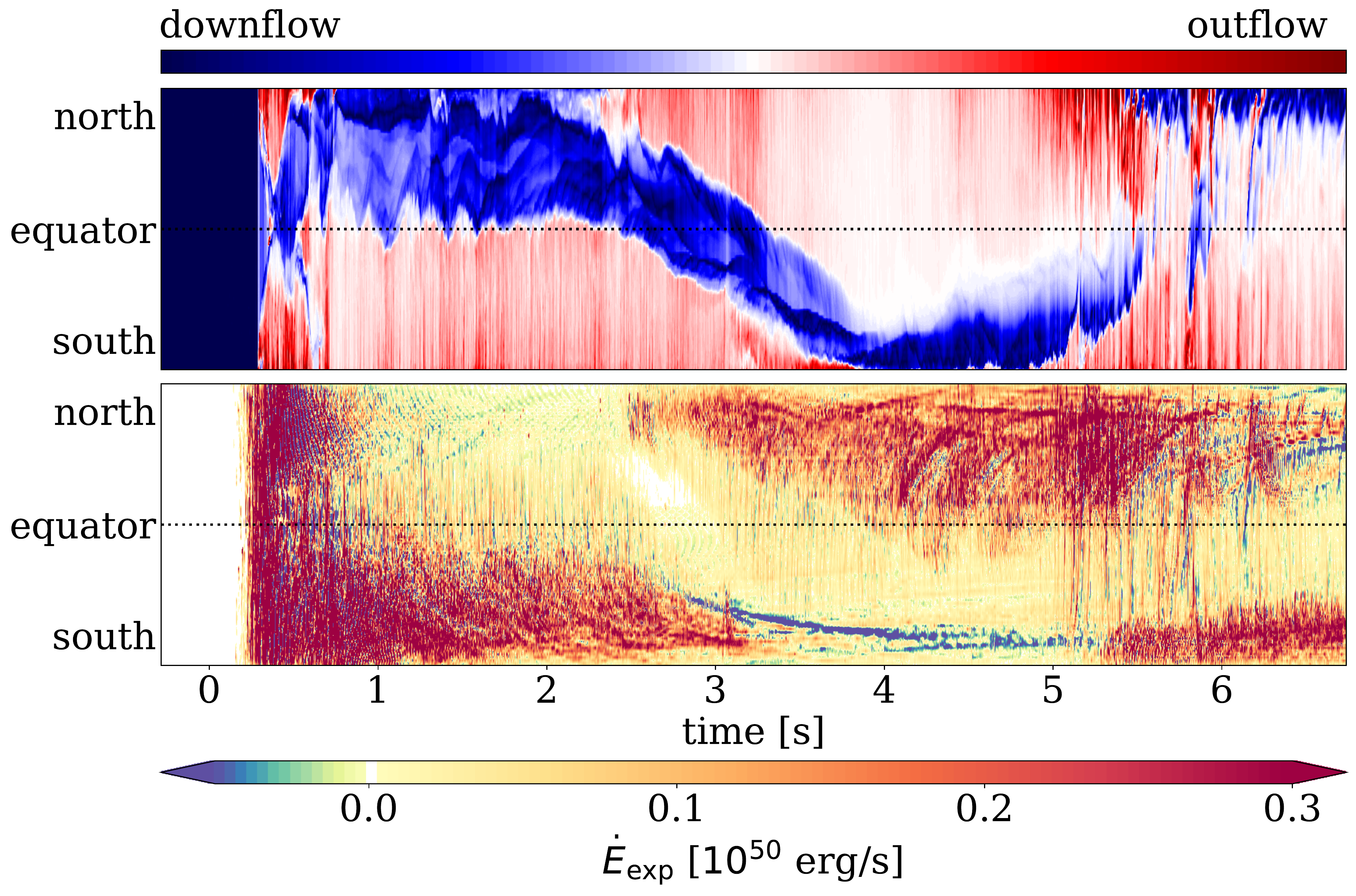}
 \caption{Directionality of downflows and explosion energy generation in model s15\_F120\_R001. The upper panel shows mass downflows and outflows (relative strength, at $r = 500$~km) in a $t$-$\theta$-plane, i.e., direction-dependent, with equator being perpendicular to the cylindrical axis. The lower panel shows the explosion energy generation rate in the same plane. The direction in which explosion energy is generated depends on the 
 direction of downflows. Time $t=0$~s corresponds to bounce.}
 \label{fig:angular_mdot_dedia_F120_R001}
\end{figure}

Downflows are a multi-dimensional, long-time, and angular dependent feature that is correlated to the explosion energy growth. We investigate the angular dependency of the downflows and explosion energy evolution defining the  explosion-energy growth rate as $\dot{E}_\mathrm{exp} \equiv \mathrm{d}E_\mathrm{exp}/\mathrm{d}t$ and comparing to the mass accretion at late times. Downflows are  derived from the mass accretion rate ($\dot{M}_\mathrm{acc}$) through a sphere with 500~km radius around the PNS. For model s15\_F120\_R001, Figure~\ref{fig:angular_mdot_dedia_F120_R001} shows the direction of downflows over time (upper panel) and how it is correlated to the explosion-energy generation (bottom panel). Before the explosion ($t \leq 0.3$~s post-bounce), there is only accretion. The explosion energy is first accumulated isotropically at shock revival. During the first seconds, the shock expands prominently towards the southern hemisphere leaving space for a long-lasting downflow from the northern hemisphere. Initially, the explosion energy grows also in directions of downflows, because we measure the accretion only at 500~km radius, but explosion energy is also accumulated below that radius. When the downflow is firmly established after $t \approx 1$~s, explosion energy is generated almost exclusively in the southern hemisphere.  The accreted matter acts as fuel for the supernova energy. The strong downflow from the northern hemisphere gradually changes its direction during $2~\mathrm{s} < t < 4~\mathrm{s}$ to the southern hemisphere and the explosion energy generation rate shifts its direction accordingly. The gravitational pull of the PNS decreases the velocity of some ejecta and leads to initially unbound matter to be bound again, which can be seen in slightly negative values of $\dot{E}_\mathrm{exp}$ in the southern hemisphere. Until the end of the simulation at $t = 6.74$~s, the downflow changes direction again and there are no signs of vanishing mass accretion yet.

\begin{figure}[ht]
 \centering
 \includegraphics[width=1.\linewidth]{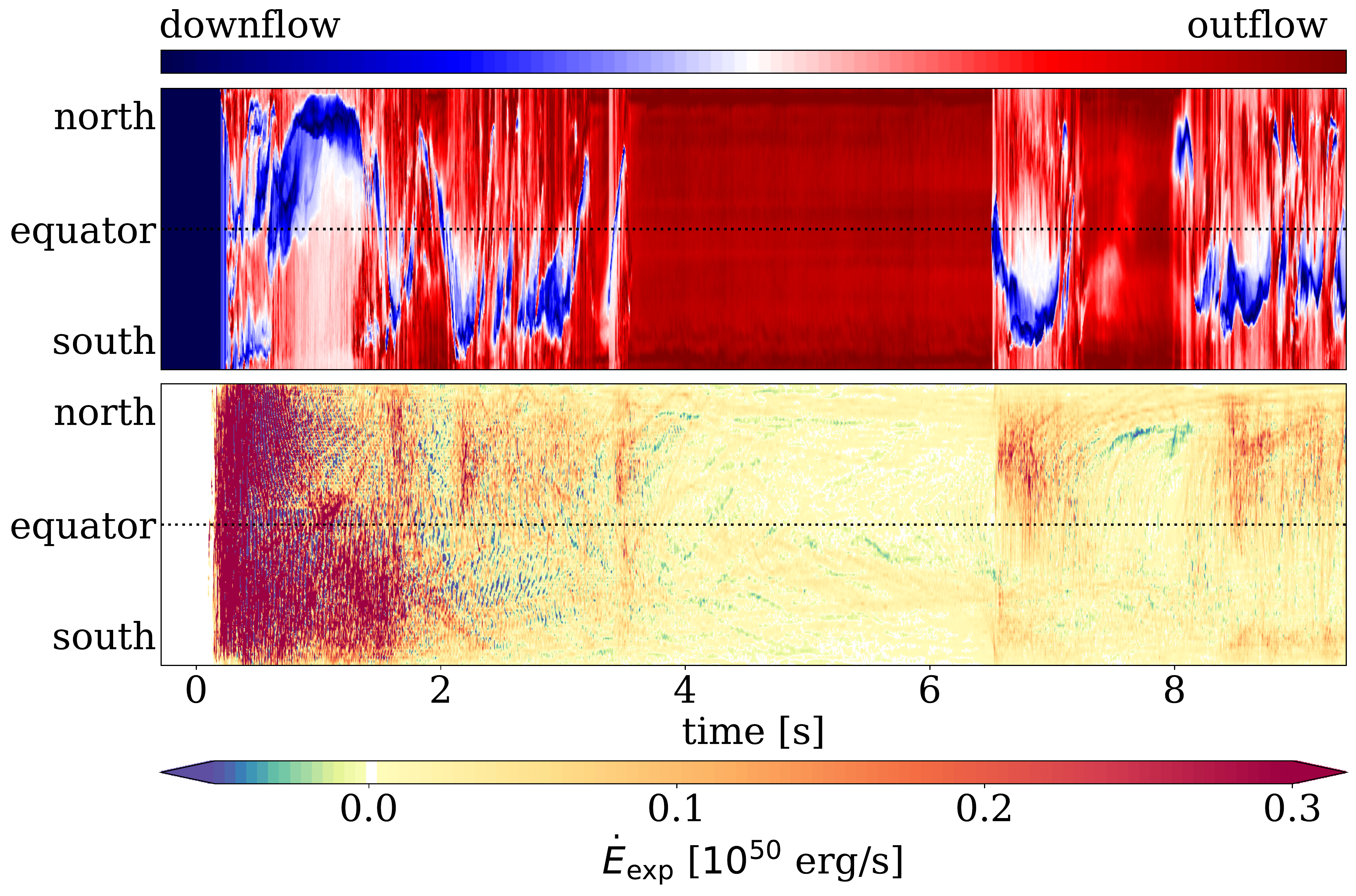}
 \caption{Same as in Fig.~\ref{fig:angular_mdot_dedia_F120_R001}, but for model s15\_F150\_R030 which has a phase of no accretion corresponding to a neutrino-driven wind (NDW).}
 \label{fig:angular_mdot_dedia_F150_R030}
\end{figure}

In some cases, we obtain explosions with less stable downflows where even a neutrino-driven wind (NDW) can form as shown in Fig.~\ref{fig:angular_mdot_dedia_F150_R030} for the model s15\_F150\_R030. The NDW phase ($3.5~\mathrm{s} < t < 6.5~\mathrm{s}$) is characterized by a vanishing mass accretion rate, $\dot{M}_\mathrm{acc} = 0$ and  matter is ejected in all directions (subsequently, we define NDW phases by having $\dot{M}_\mathrm{acc} = 0$ for at least $10$~ms duration). During this wind phase, considerably less explosion energy  is added. However, $\dot{E}_\mathrm{exp}$ does not vanish completely, since there is a continuous outflow of matter ejected by neutrinos when depositing energy in the layers around the PNS. At $t \approx 6.5$~s, the wind is terminated by a downflow, and consequently a new phase of explosion energy-generation sets in. This transition occurs almost instantaneously. For all models that develop a NDW (see Sect.~\ref{sec:long-time_observables}), we see typical integrated values of $\dot{E}_\mathrm{exp} \approx 0.04$~B~s$^{-1}$ during phases of no accretion.  Our results suggest that this ongoing interplay of accretion and ejection can continue for longer than previously thought. However, 3D simulations would be necessary to conclude the impact and duration of downflows and NDWs.

\subsection{Impact of rotation and explosion morphology}
\label{sec:long-time_dshock}
\begin{figure}[ht]
 \centering
 \includegraphics[width=1.0\linewidth]{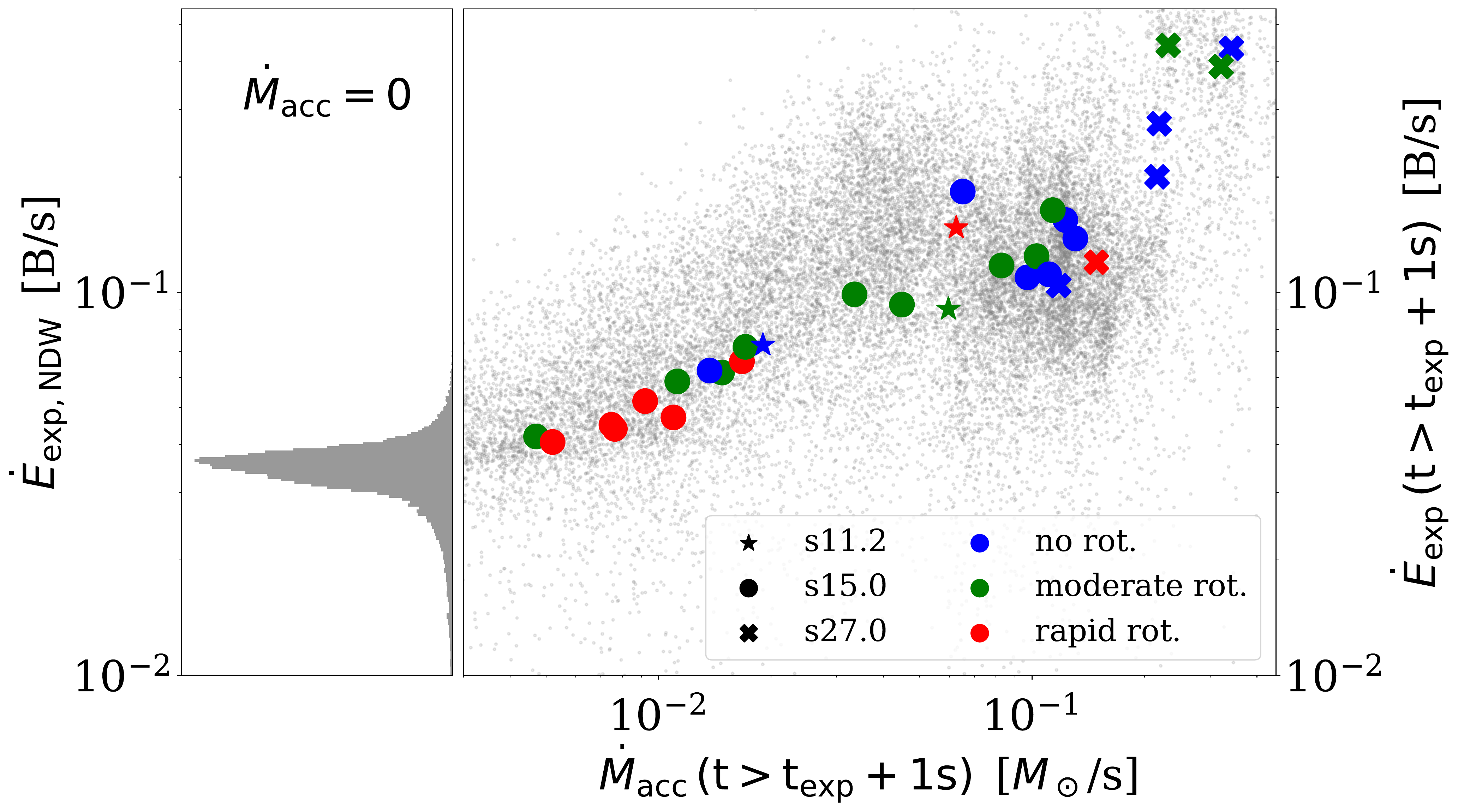}
 \caption{Right panel: Growth-rate of the explosion energy versus mass accretion at late times (i.e.,$t>t_\mathrm{exp}+1$~s). The gray dots represent individual data points (shown as running averages of 5~ms length) from all simulations during the long-time evolution, while the colored symbols indicate the mean values from different models for that phase.  Left panel: Density distribution of $\dot{E}_\mathrm{exp}$ during NDW phases with $\dot{M}_\mathrm{acc} = 0$.}
 \label{fig:scatter_dedia_vs_mdot}
\end{figure}

The generation of explosion energy through mass accretion at late times is a robust mechanism in our models. At one second after the explosion (i.e., at the start of the long-time phase), the initial shock wave has reached a radius of $(1-2) \cdot 10^{4}$~km, has cooled down to $T\sim 0.5-1.5$~GK, and has reached densities of $\rho \sim 10^5~\mathrm{g~cm^{-3}}$. At this point, the explosion energy stored within the shock wave and the ejecta generated earlier is approximately constant and any additional $E_\mathrm{exp}$ generation originates from the vicinity of the PNS. Therefore, there is a correlation between $\dot{M}_\mathrm{acc}$ and  $\dot{E}_\mathrm{exp}$ for all exploding models as shown in Fig.~\ref{fig:scatter_dedia_vs_mdot}. In this figure both quantities are shown for the long-time evolution, i.e. $t > t_\mathrm{exp} + 1$~s. The gray dots correspond to individual times (each dot is a mean value over a 5~ms time interval) of all simulations and the colored symbols are obtained by averaging during the whole long-time phase. This introduces some bias for models with  short simulation times because the mass accretion is higher during the the first seconds. In any case,  there is a  correlation  of late-time accretion and explosion-energy growth rate.
In NDW phases of no accretion, the explosion energy growth rate adopts values of $\dot{E}_\mathrm{exp,NDW} = (0.035 \pm 0.007)~\mathrm{B~s^{-1}}$ (left panel of Fig.~\ref{fig:scatter_dedia_vs_mdot}). We note that this specific value might be influenced by our use of a gray neutrino leakage scheme and Newtonian gravity.

The early shock morphology has a clear impact on the formation of stable downflows and thus on the late evolution of accretion. In order to quantify the shock morphology, we employ the shock deformation parameter introduced by \citet{Scheck:2006}. It uses the shock radius $R_\mathrm{s}$ as a function of the polar angle $\theta$ and is given by
\begin{equation}
 \label{eq:dshock}
 d_\mathrm{shock} \, = \, \frac{ \mathrm{max} [ R_\mathrm{s}(\theta) \, \mathrm{cos}(\theta) ] - \mathrm{min} [ R_\mathrm{s}(\theta) \, \mathrm{cos}(\theta) ] }
                    {2 \cdot \mathrm{max} [ R_\mathrm{s}(\theta) \, \mathrm{sin}(\theta) ]} - 1 \, .
\end{equation}
The parameter is equivalent to the ratio of the maximum shock diameters, parallel and perpendicular to the cylindrical axis. It can have positive and negative values, for a prolate and an oblate shock deformation, respectively. In the case of a spherical shock expansion, $d_\mathrm{shock}$ becomes zero. We find a  correlation between the shock deformation parameter at the time of shock revival and the rotation strength   (see Tab.~\ref{tab:model_overview}).  Non-rotating 2D simulations typically explode in a prolate morphology, which is a known feature of this geometry \citep[see e.g.,][]{Mueller:2015_2, Nakamura:2015, Bruenn:2016, Summa:2016, OConnor:2018_1, Vartanyan:2018}. However, the resulting accretion along preferred directions has also been observed in 3D simulations \citep{Burrows:2019,Vartanyan:2019_1}. 

Rapidly rotating models tend to have less accretion due to  the increased centrifugal forces, which reduce the infall velocity of matter from the equatorial plane and therefore also the accretion rate. The impact of rotation is present in all phases of a CCSN. With increased rotation rate, the shock morphology becomes more spherical and eventually also slightly oblate, for rapidly rotating models. An exception here are simulations of the s11.2 progenitor, due to the smaller amount of angular momentum  (a factor of $10$ less than s15.0 models with the same rotation rate) and an earlier explosion time, which allows for less angular momentum to be accreted until shock revival.

\begin{figure}[ht]
 \centering
 \includegraphics[width=1.0\linewidth]{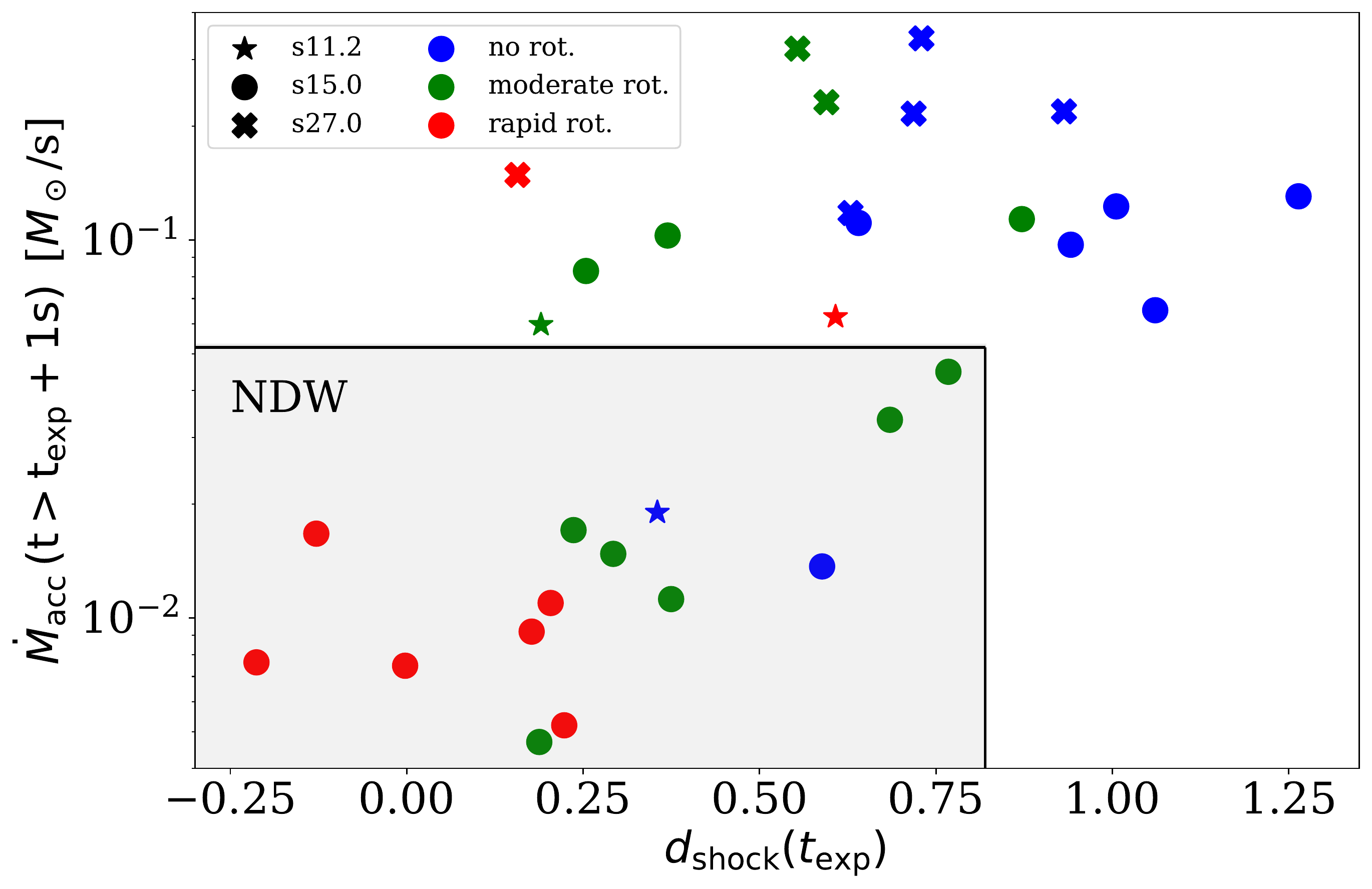}
 \caption{Average late-time accretion rate versus the shock deformation parameter at explosion time for the three progenitors (different symbols) and various rotation strengths (different colours). Models within the shaded area have a phase of zero accretion or NDW during the longtime evolution.}
 \label{fig:scatter_mdot_vs_dshock}
\end{figure}

We can investigate the relation between shock deformation (Eq.~\ref{eq:dshock}) and late mass accretion. Downflows usually originate from directions with moderate early shock expantion. In those directions,  part of the ejected matter  does not reach the escape velocity and eventually falls back onto the PNS. Figure~\ref{fig:scatter_mdot_vs_dshock} shows the relation of the early shock morphology $d_\mathrm{shock}(t_\mathrm{exp})$ and the mass accretion at late times. There is not a clear correlation of these two quantities but some trends. Simulations exploding with $d_\mathrm{shock}(t_\mathrm{exp}) \geq 0.5$ are typically moderately rotating or non-rotating (with the exception of the model s11\_F100\_R020) and consistently have an average mass accretion at late times of $\dot{M}_\mathrm{acc} > 0.01~M_\odot/\mathrm{s}$. Furthermore, we can separate all simulations in two groups depending on whether they have some phase of zero accretion. The models that develop a NDW are in the lower left corner region corresponding to lower late accretion and not extreme shock deformation.

\subsection{Evolution in the long-time phase and the neutrino-driven wind}
\label{sec:long-time_observables}

As shown in previous section, there is a strong link between the shock deformation and the formation of stable downflows that critically affect the explosion energy. Here, we want to study the long-time evolution of the explosion energy and mass ejection. For the s15 progenitor, we compare  these quantities at two different times, $1$~s and $5$~s after shock revival.

\begin{figure}[ht]
 \centering
 \includegraphics[width=0.9\linewidth]{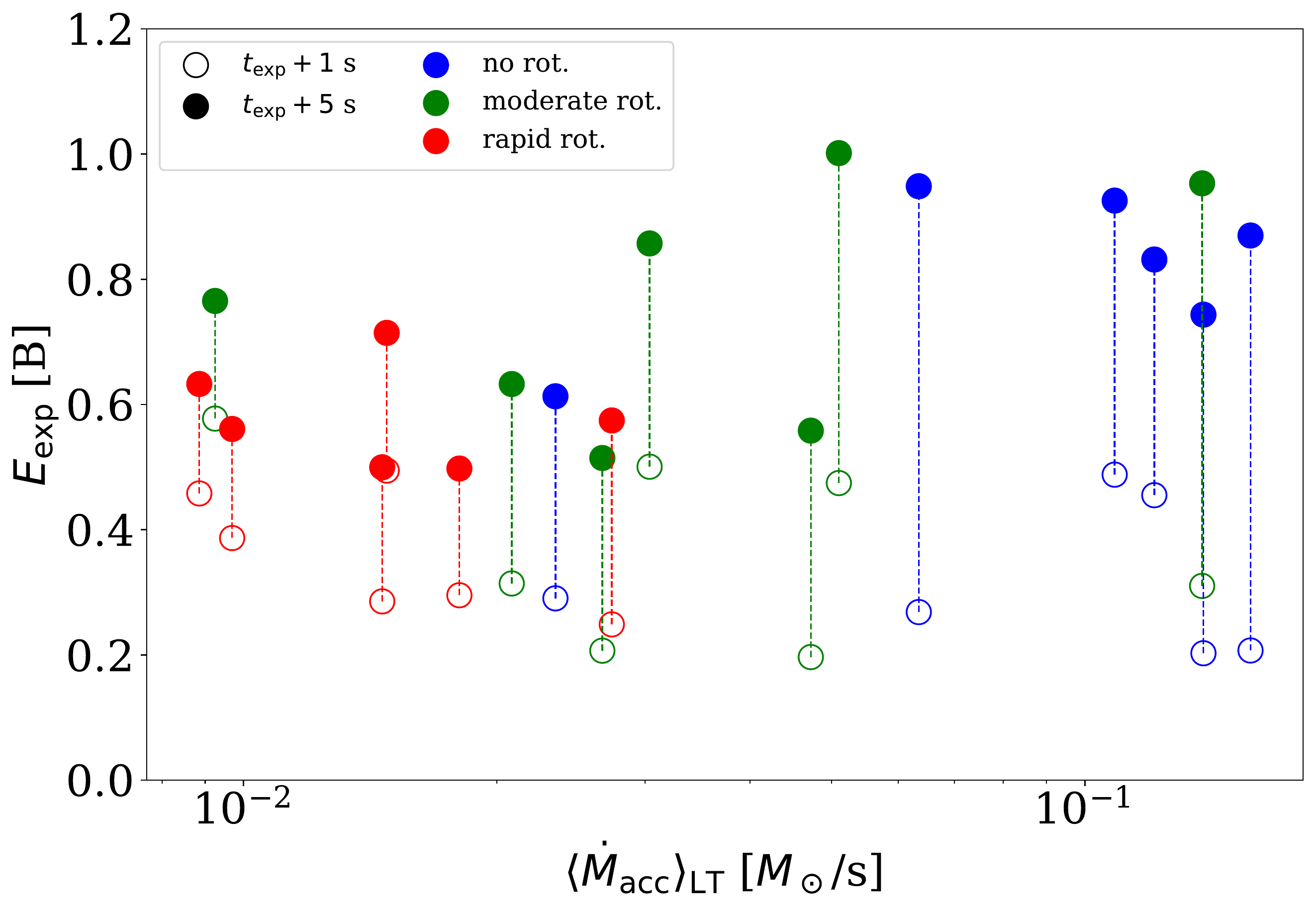}\\
 \includegraphics[width=0.9\linewidth]{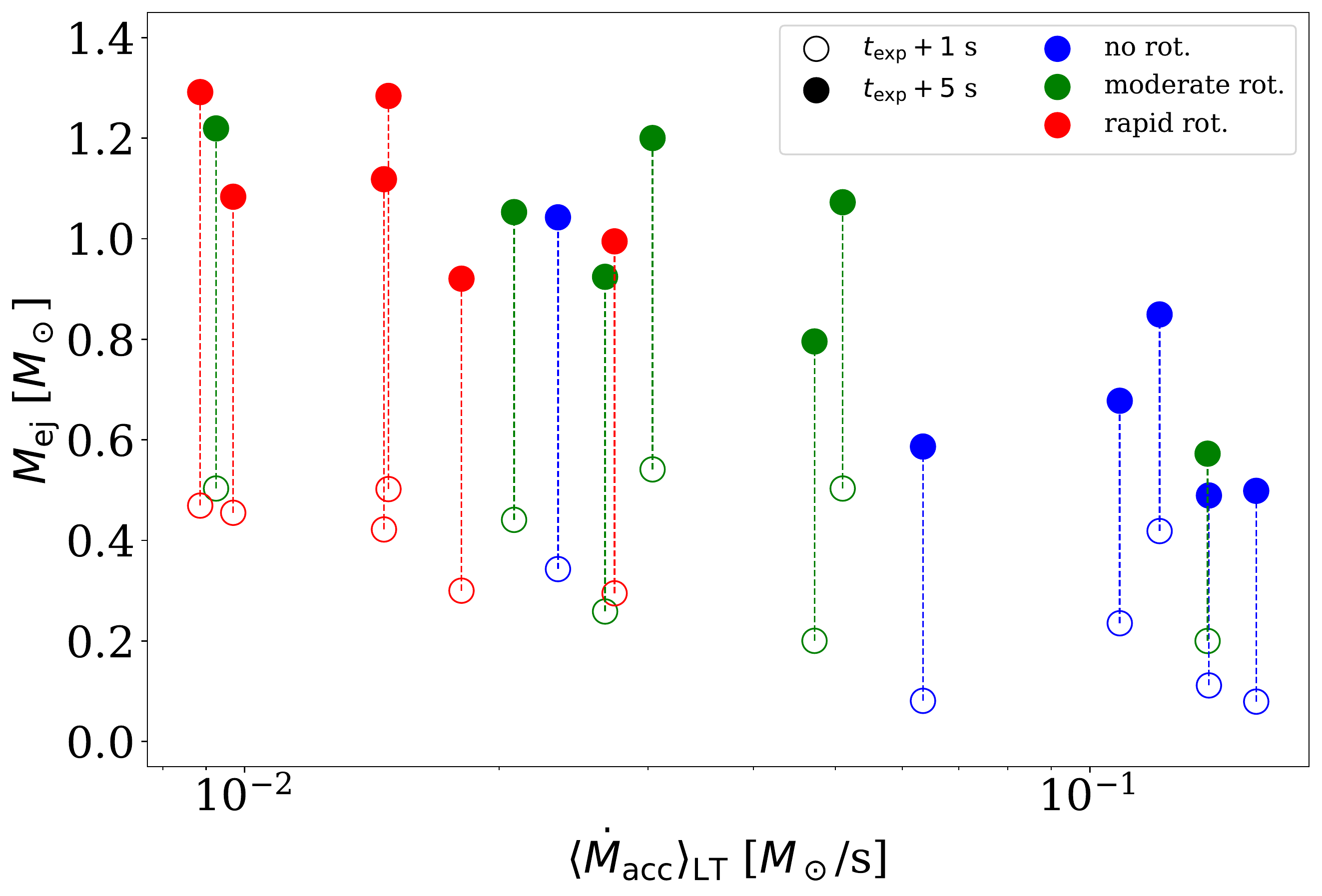}
 \caption{Influence of the long-time evolution on the explosion energy (upper panel) and mass ejection (bottom panel) related to the mass accretion. Colors indicate the rotation strength. Open and close circles show the explosion energy at $1$~s and $5$~s after shock revival, respectively. The mass accretion is averaged   for the time interval $1 < t - t_\mathrm{exp} < 5$~s denoted as LT for long-time.}
 \label{fig:change_1_5s}
\end{figure}

The evolution of the explosion energy and its dependence on the mass accretion is shown in Fig.~\ref{fig:change_1_5s}, upper panel. At $1$~s after shock revival, the explosion energy of all models is  distributed at $(0.4 \pm 0.2)$~B, independent of the rotation strength and mass accretion. During the next $4$~s, rapidly rotating models increase their explosion energy by  $\sim 0.4$~B and end up with  values below $0.8$~B at $t_\mathrm{exp} + 5$~s. In contrast,  moderately rotating  and non-rotating models gain considerably more energy during this same time reaching around 0.9-1~B. The value for non-rotating models is also in good agreement with recent 3D studies \citep{Bollig:2020}. We note that for simulations that run for a longer time, the final explosion energy keeps increasing after $t_\mathrm{exp} + 5$~s and the trend with rotation is consistent also at later times. Therefore, we conclude that, for models with initial morphology favourable for large-scale downflows,  the saturation point of the explosion energy lies beyond $10$~s after the initial explosion.
\begin{figure}[ht]
 \centering
 \includegraphics[width=1.0\linewidth]{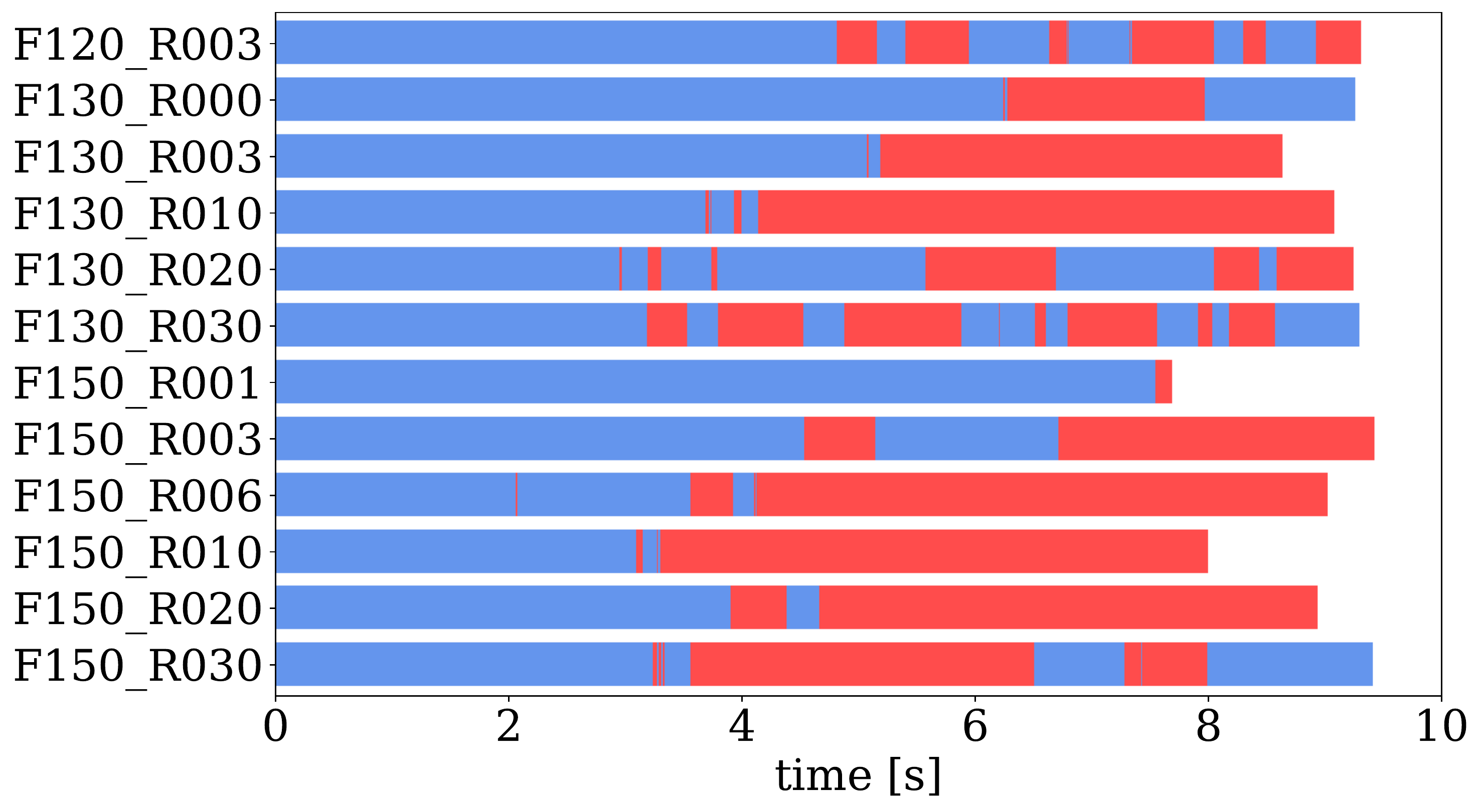}
 \caption{Timelines of s15.0 progenitor simulations, where red areas mark the presence of an isotropic neutrino-driven wind phase. We omit the ``s15\_'' prefix in the model names.}
 \label{fig:wind_duration}
\end{figure}

The long-time evolution is also critical for nucleosynthesis (Sect.~\ref{sec:nucleosynthesis}), therefore here we investigate  the amount of mass ejected (Fig.~\ref{fig:change_1_5s}, bottom panel). At $1$~s after shock revival, models with lower accretion (corresponding to approximately spherical explosions, Fig.~\ref{fig:scatter_mdot_vs_dshock}) have ejected more mass than those with strong downflows and a prolate morphology. This is due to the more energetic shock expansion into the equatorial direction in spherical models compared to prolate explosions. This trend continues also in the next seconds. We calculate the average energy of ejected matter: For low accretion models (spherical explosions), we obtain values of $E_\mathrm{exp}/M_\mathrm{ej} \approx 0.5~\mathrm{B}/M_\odot$, while models with high accretion and prolate morphologies reach $E_\mathrm{exp}/M_\mathrm{ej} \approx 1.5~\mathrm{B}/M_\odot$ at $5$~s.

\begin{figure*}[!ht]
 \centering
 \includegraphics[width=0.8\linewidth]{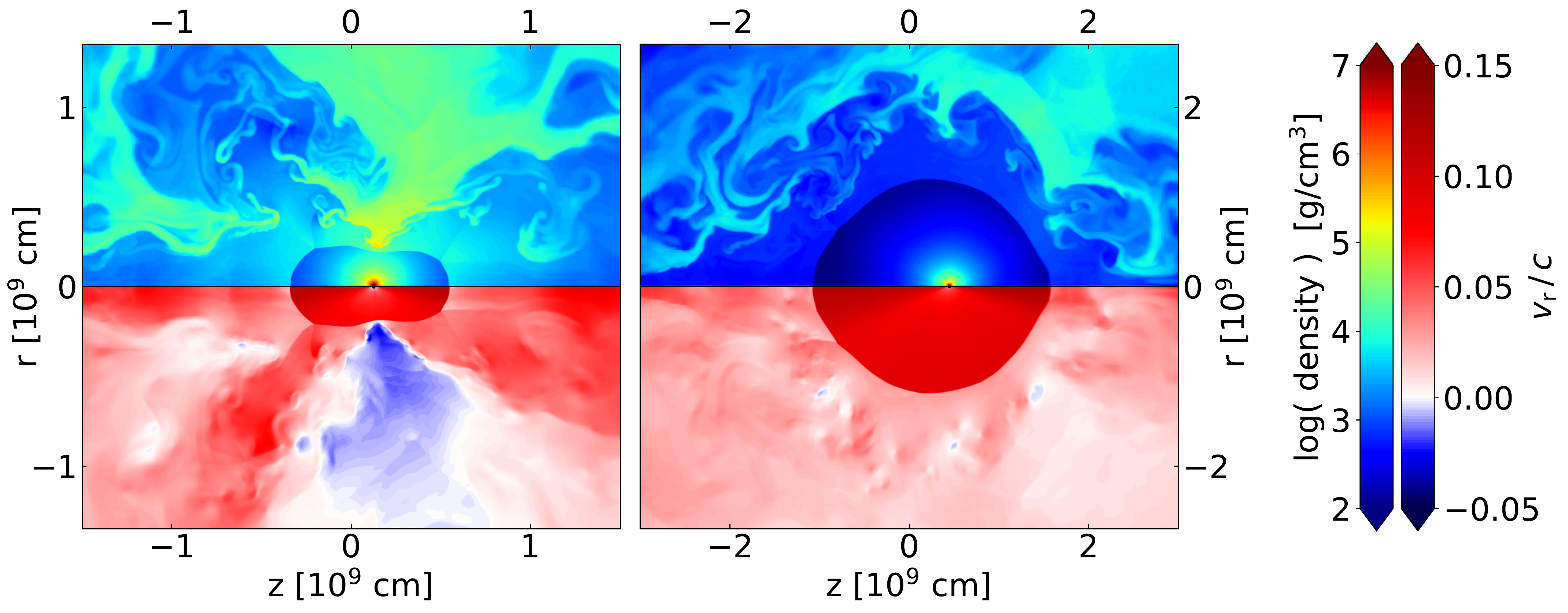}
 \caption{Density (top) and radial velocity (bottom) of model s15\_F150\_R030 (left) at 6.3~s shortly before termination of the wind (compare to Fig.~\ref{fig:angular_mdot_dedia_F150_R030}~\&~\ref{fig:wind_duration}), and model s15\_F150\_R020 (right) at 8~s after bounce. Note the different axes scales for both models.}
 \label{fig:wind_2D_twoPanel}
\end{figure*}
Even if a considerable amount of mass is ejected in all exploding models, a neutrino-driven wind develops only in 12 of the 21 exploding models for the s15 progenitor. These isotropic NDW phases ($\dot{M}_\mathrm{acc} = 0$ for at least $10$~ms duration) are shown in red in Fig.~\ref{fig:wind_duration}. The first wind phases start to appear after $t \approx 3$~s. There are two kinds of NDW phases: short- and long-duration. The termination of a NDW is usually due to large accumulations of matter with negative radial velocity above the wind. We show examples of the two possible NDW phases in Fig.~\ref{fig:wind_2D_twoPanel}. Here, the symmetry axis is displayed horizontally and we show the density in the upper half of the domain, and the radial velocity in its lower half. The NDW regions are visible around the high-density region with the PNS in the center, and are characterized by high, supersonic velocities, a steep density gradient and the wind termination shock, where the density (velocity) increases (decreases) abruptly. In the example of model s15\_F150\_R030, one can see the matter accumulations with negative radial velocity and comparably high density, just above the wind termination shock. Long-duration winds can extend up to several $10^4$~km radius (see model s15\_F150\_R020 in Fig.~\ref{fig:wind_2D_twoPanel}). The duration of the wind depends on the heating factor and rotation rate. When increasing the heating factor and/or the rotation, there is lower mass accretion  and this allows a wind to form and last for several seconds. Other 2D and 3D simulations indicate also that there is not always a neutrino-driven wind after a successful explosion \citep[see, e.g., ][]{Bruenn:2016,OConnor:2018_1,Vartanyan:2019_1,Bollig:2020}.

\section{Nucleosynthesis}
\label{sec:nucleosynthesis}

Since our neutrino treatment is very simple, the following nucleosynthesis results are only approximate, even if we correct the neutrino properties affecting the evolution of the electron fraction. We focus only on the 21 exploding simulations of the 15~$M_\odot$ progenitor (see Table~\ref{tab:model_overview}). The amount of ejected mass  (Fig.~\ref{fig:change_1_5s}) corresponds to several thousands of ejected tracers per model. These ejected tracers can be divided into ``neutrino-processed'' and ``shock-processed''. Tracers are considered neutrino-processed, when their electron fraction changes by $|\Delta Y_\mathrm{e}| > 0.01$ compared to their starting value. Most of these tracers reach rather high temperature and enter NSE resetting the progenitor composition. Shock-processed particles retain their original electron fraction, their temperature is high enough to change slightly the progenitor composition but most of them stay too cold to enter NSE.

\begin{figure}[!ht]
 \centering
 \includegraphics[width=\linewidth]{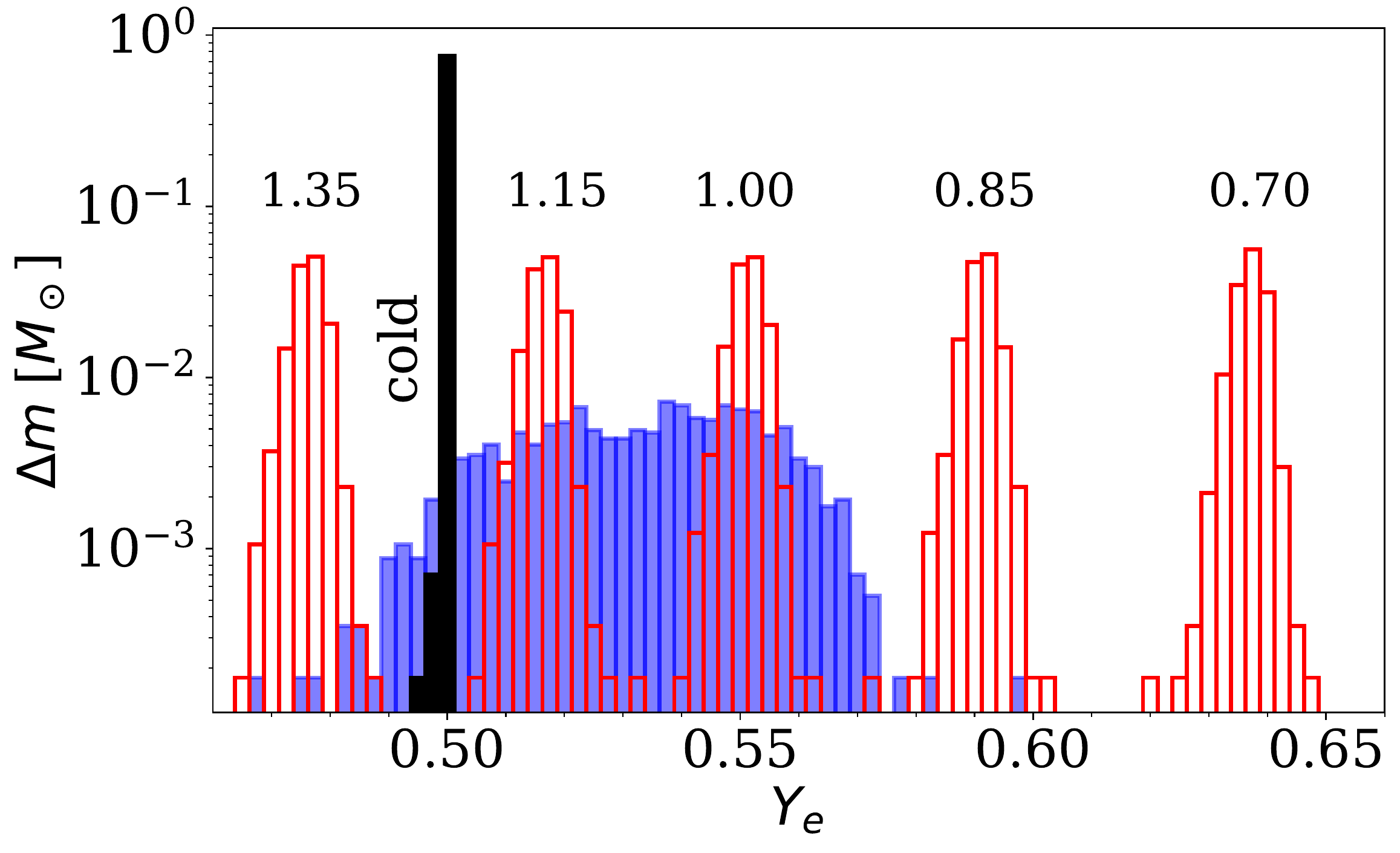}
 \caption{Modified  distributions for the initial $Y_e$ of model s15\_F120\_R006 when the trajectory cools down to $10$~GK, or at their peak temperature for cold tracers. The black filled bars correspond to the cold (shock-processed) tracers for all distributions. The blue bars show the distribution with the original neutrino energies. The red bars show the distributions with fixed neutrino energy difference and fixed number luminosity ratios, which are indicated above.}
 \label{fig:F120_R006_allYeDist}
\end{figure}

For the cold, shock-processed tracers that do not reach NSE, the evolution of $Y_e$ is not affected by neutrinos.  Therefore, for these tracers we use the electron fraction obtained in the simulations. In contrast, the hot tracers are sensitive to neutrino quantities that determine $Y_e$, namely neutrino and antineutrino energies and number luminosities \citep{Qian:1996}. Therefore, for those tracers we try different corrections to cover all possible conditions. We investigate two different corrections to the $Y_e$ based on one model and later extend our study to the remaining 20 explosions of the 15~$M_\odot$ progenitor. We select model s15\_F120\_R006 as reference here because it is the closest to the ``s15'' model in \citet{Wanajo:2018} when comparing $t_\mathrm{exp}$, $E_\mathrm{exp}$, $M_\mathrm{PNS}$ and $M_\mathrm{ejected}$. However, our neutrino energies and luminosities based on the neutrino leakage scheme lead to unrealistically neutron-rich conditions.

\begin{figure*}[!ht]
 \centering
  \includegraphics[width=0.9\linewidth]{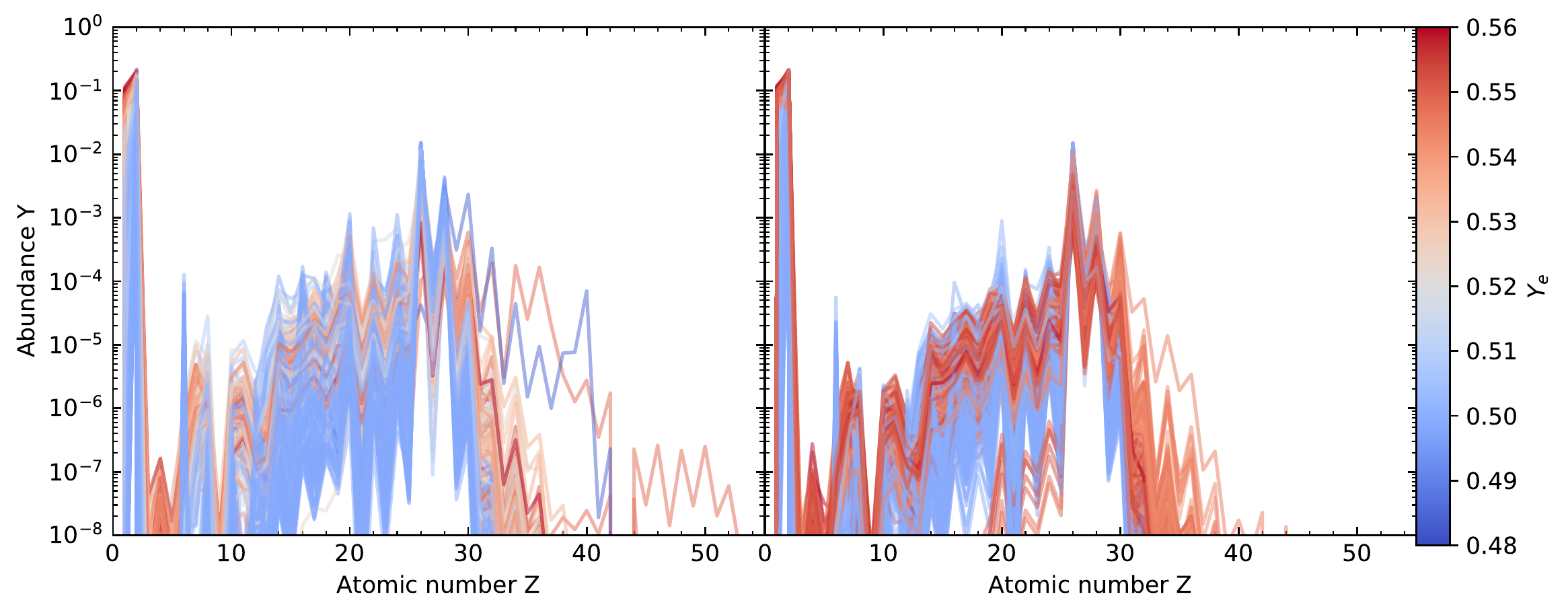}
 \caption{Abundances of all ejected neutrino-processed tracers of model s15\_F120\_R006 based on broad (left) and narrow with $RL_{\mathrm{n}} = L_{\mathrm{n},\bar{\nu}_e} / L_{\mathrm{n},\nu_e}=1$ (right) $Y_e$ distributions  shown in  Fig.~\ref{fig:F120_R006_allYeDist}. The colours of the lines indicate the electron fraction of the individual tracers at 5.8~GK.}
 \label{fig:ab_F120R006}
\end{figure*}

In the determination of the electron fraction, there are two critical quantities: the energy difference between antineutrinos and neutrinos ($\Delta \epsilon_\nu =  \langle\epsilon_{\bar{\nu}_e}\rangle - \langle\epsilon_{\nu_e}\rangle$) and the ratio of neutrino number luminosities ($RL_{\mathrm{n}} = L_{\mathrm{n},\bar{\nu}_e} / L_{\mathrm{n},\nu_e}$). Following previous studies based on simulations with accurate neutrino transport  \cite[see e.g.,][]{Liebendoerfer:2005_2,Bruenn:2016,Takiwaki:2016,Kotake:2018,Cabezon:2018,Just:2018,OConnor:2018_1,Summa:2018,Vartanyan:2018,Vartanyan:2019_2,Mueller:2019_2,Pan:2019,Powell:2020,Kuroda:2020}, one can find typical energy differences of $\Delta \epsilon_\nu =  \langle\epsilon_{\bar{\nu}_e}\rangle - \langle\epsilon_{\nu_e}\rangle \approx 2 - 2.5$~MeV. In our models we have on average $\Delta \epsilon_\nu = (2.3 \pm 0.6)$~MeV after the explosion time.

In our first approach, we use the original neutrino energies from our simulation and  correct the luminosities similarly to \citet{Sieverding:2020}, but adopting a constant luminosity ratio of $L_\mathrm{\mathrm{n},\bar{\nu}_e} /L_{\mathrm{n},\nu_e}= 1.25$ instead of $1$. This results in a $Y_e$ distribution of ejected matter  that is comparable to the studies of \citet{Wanajo:2018}  and \citet{Sieverding:2020}. The blue histogram in Fig.~\ref{fig:F120_R006_allYeDist} shows the corrected electron fraction distribution, with the cold tracers as black filled bars. The distributions for the hot tracers correspond to the initial electron fraction in the network calculations and corresponds to a temperature of $\sim 10$~GK. For the hot tracers, the  abundances obtained based on this distribution are shown in Fig.~\ref{fig:ab_F120R006} (left panel) where each line corresponds to an individual tracer and the colors indicate the $Y_e$ value at 5.8~GK, i.e. around the temperature when the approximation of NSE breaks down\footnote{Notice that between the initial $Y_e$ at  $\sim 10$~GK shown in Fig.~\ref{fig:F120_R006_allYeDist} and the $Y_e$ at 5.8~GK (Fig.~\ref{fig:ab_F120R006}), the network assumes NSE but the weak reactions are still evolved leading to an evolution of the electron fraction.}.  Iron-group nuclei dominate the final abundances and few tracers at the extremes of the $Y_e$ distribution reach the region of Sr, Y, and Zr.

In our second approach to correct the $Y_e$, we change our neutrino energies to match more closely the literature values. We subtract $4$~MeV from the $\bar{\nu}_e$ leakage energy and adopt a constant energy difference of $\Delta \epsilon_\nu =  \langle\epsilon_{\bar{\nu}_e}\rangle - \langle\epsilon_{\nu_e}\rangle = 2$~MeV. Moreover, we parametrize the ratio of number luminosities, $RL_\mathrm{n} = L_{\mathrm{n},\bar{\nu}_e} / L_{\mathrm{n},\nu_e} = 0.70, 0.85, 1.00, 1.15, 1.35$, where a smaller  ratio corresponds to more proton-rich conditions. This correction leads to very narrow $Y_e$ distributions as shown in Fig.~\ref{fig:F120_R006_allYeDist}. These distributions disagree with those from state-of-the-art simulations, however they are very useful to understand how the abundances depend on a given electron fraction in the supernova ejecta and to cover all possible conditions. Figure~\ref{fig:ab_F120R006} (right panel) shows the abundances for the hot tracers of model s15\_F120\_R006 based on the narrow distribution with $RL_\mathrm{n} = L_{\mathrm{n},\bar{\nu}_e} / L_{\mathrm{n},\nu_e} =1$. In that case as well, the majority of the produced nuclei lie in the iron peak, with a small portion of tracers producing nuclei with $Z>30$.

\begin{figure}[ht]
 \centering
 \includegraphics[width=1.0\linewidth]{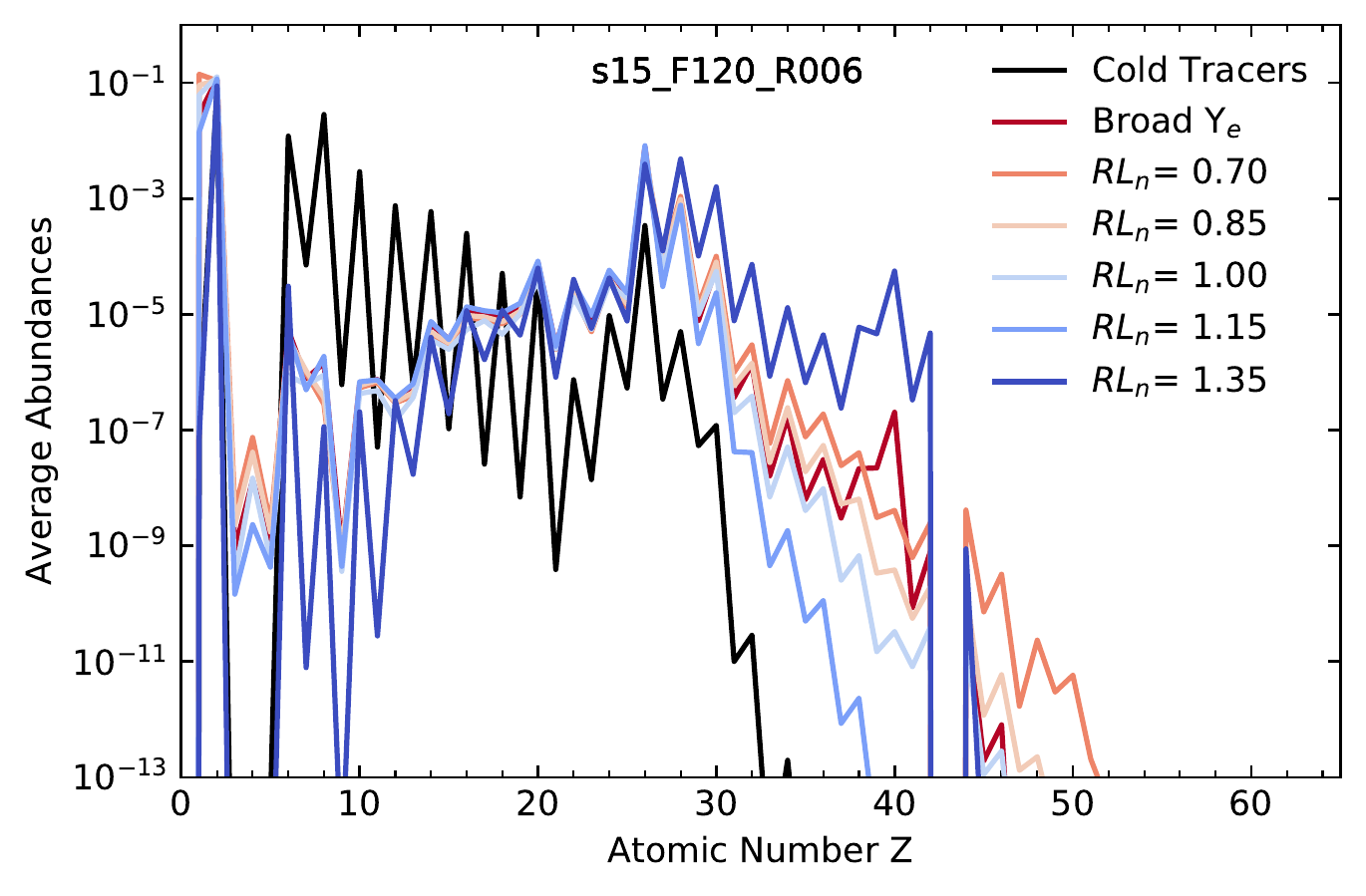}
 \caption{ Integrated abundances of all trajectories of  model s15\_F120\_R006 and  different $Y_e$ distributions as given in Fig.~\ref{fig:F120_R006_allYeDist}.}
\label{fig:ab_overview_F120_R006}
\end{figure}

\begin{figure}[ht]
 \centering
  \includegraphics[width=1.0\linewidth]{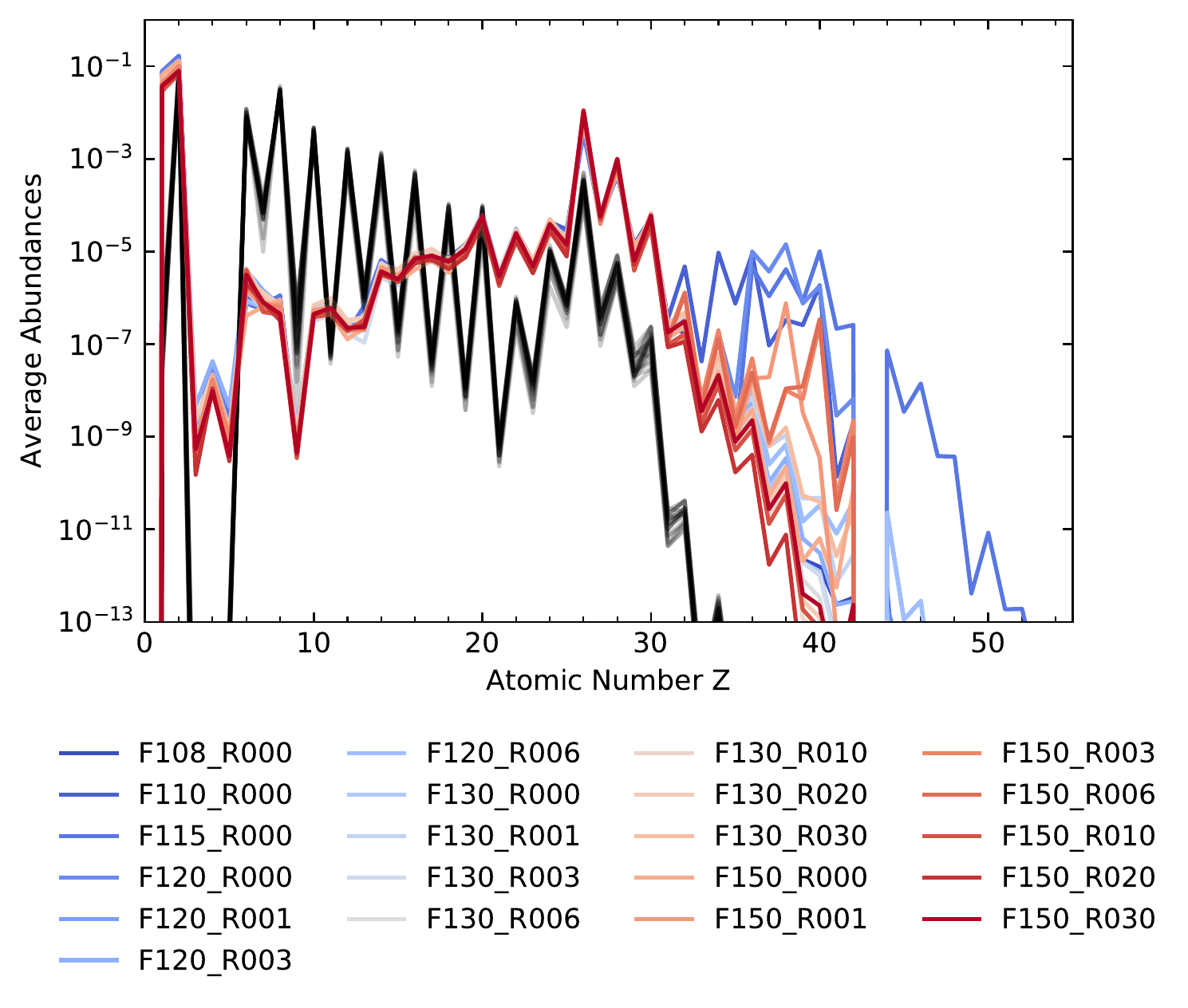}
 \caption{ Integrated abundances of all trajectories for all exploding models of s15 and the narrow $Y_e$ distributions with $RL_\mathrm{n}=1$. The black lines represent the abundances from the shock--heated tracers from the different models.}
\label{fig:ab_overview}
\end{figure}

\begin{figure}[ht]
 \centering
 \includegraphics[width=1.0\linewidth]{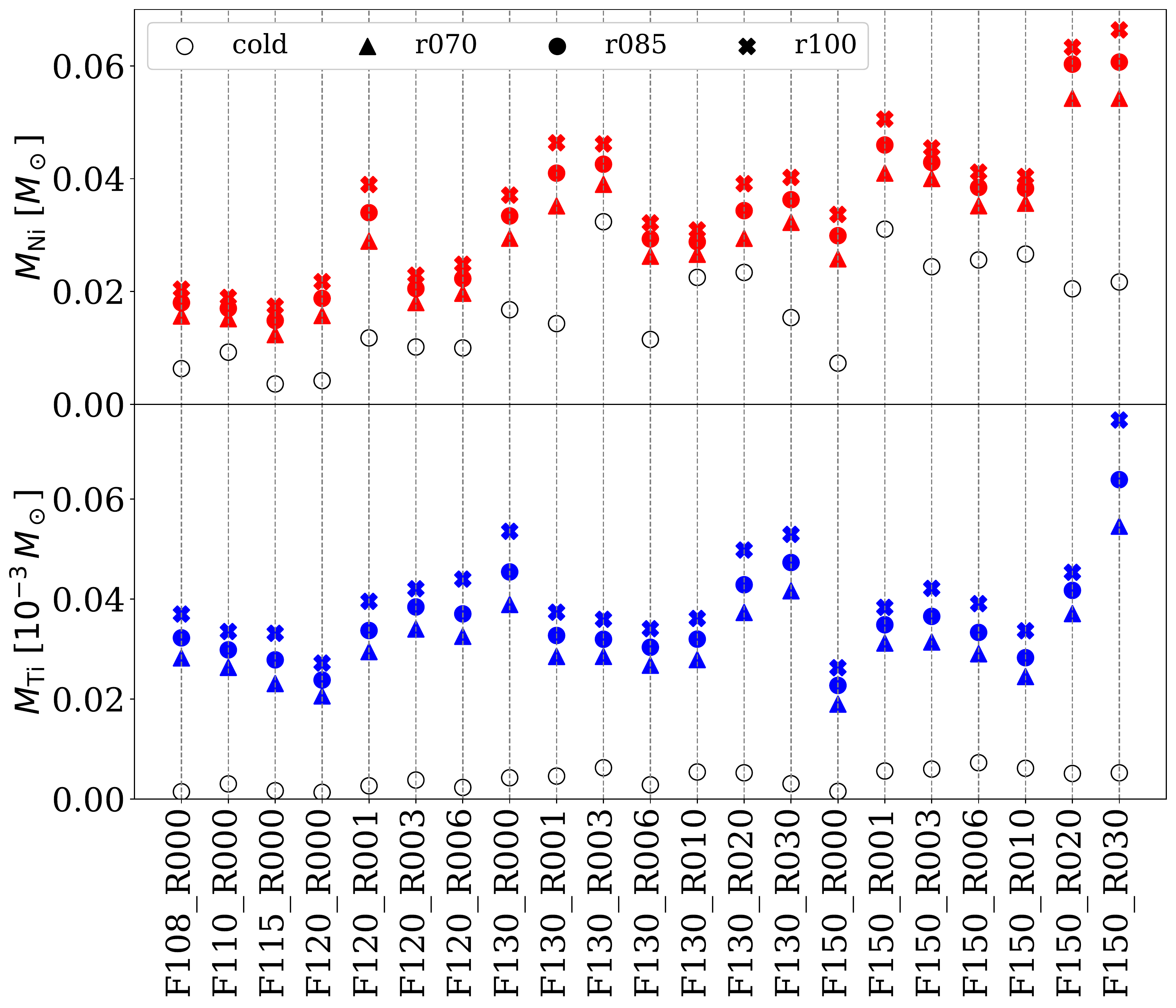}
 \caption{Yields of $^{56}$Ni, $^{44}$Ti obtained by post-processing the 21 exploding models. For each model, the yields are shown for the cold tracers (empty circles) and the neutrino-processed, hot tracers (filled symbols) for three different $Y_e$ narrow distributions corresponding to $RL_\mathrm{n}=$0.7, 0.85, and 1 (triangles, circles, and crosses, respectively). }
\label{fig:all_Ni_Ti}
\end{figure}

The abundances for cold  and hot tracers with different $Y_e$ treatment are shown in Fig.~\ref{fig:ab_overview_F120_R006} for model s15\_F120\_R006. In general, the abundances for cold and hot tracers show a clear iron peak. The cold component is characterized by the odd-even distribution from carbon to calcium, following the progenitor composition. The hot component reaches heavier elements than the cold one with a slight dependency of the abundances on the exact electron fraction.  The different assumptions for the initial $Y_e$ lead to some variability for the abundances beyond iron covering all expected possibilities for yields from neutrino-driven supernovae. The production of elements in the region of Sr, Y, Zr is more efficient for slightly neutron-rich conditions, corresponding to $RL_{\mathrm{n}} = 1.35$ \citep{Arcones:2014, Arcones.Montes:2011}. We observe a very similar behavior  and dependency of abundances on the $Y_e$ distribution for all  exploding models of the 15~$M_\odot$ progenitor. As a summary, we show in Fig.~\ref{fig:ab_overview}  the final abundances only for the narrow $Y_e$ distribution with $RL_\mathrm{n}=1$ for all models. Due to our necessary but artificial correction of the electron fraction, we do not find any clear correlation between the abundances and the explosion energy, accretion, and rotation. Simulations with detailed neutrino transport are necessary to narrow the uncertainties in the abundances and to link those to other astrophysical conditions. Even if our results are not completely conclusive for abundances beyond iron, they indicate that there is not a strong variability for iron group elements. This may be important to estimate uncertainties in the production of $^{56}$Ni and $^{44}$Ti as shown in Fig.~\ref{fig:all_Ni_Ti} where we present an overview of all models including cold and hot components and variations of $Y_e$.

\section{Conclusions}
\label{sec:conclusions}

We have presented a broad study of long-time effects based on two-dimensional simulations of CCSN. We define long-time as starting one second after the explosion and following the evolution up to around 10~s post bounce. Our study is based on three  progenitors and variations of neutrino heating  and rotation rates. In total, we present 46 models with 32 of them exploding. In order to be able to run so many simulations for several seconds, we have used for the neutrinos a simple leakage scheme. We are aware that the 2D and neutrino treatment simplifications are critical to make any quantitative conclusion. However, we have found interesting features and trends that we expect to be present in 3D, detailed neutrino transport simulations. Our results indicate that the evolution during the first seconds after shock revival is not negligible, and that accretion and long-lasting downflows impact the growth of the explosion energy. The evolution during the long-time phase can be linked to initial conditions such as the rotation rate and the shock deformation. We find that neutrino-driven winds (NDWs) appear favorably in models with increased neutrino heating and rotation, and can be either long lasting (stable) or short lived (unstable). Improved simulations following the evolution  after the explosion are required to understand CCSN and the connection to observables.

During the explosion phase, both rotation and increased neutrino heating impact the shock acceleration and the initial explosion energy. Rotation weakens the explosion by decreasing the mass accretion and resulting accretion luminosity, which agrees with previous studies \citep{Fryer:2000, Kotake:2003, Buras:2003, Thompson:2004, Marek:2009,Summa:2018,Obergaulinger:2020}. On the other hand, an increased heating factor typically leads to more powerful explosions \citep{OConnor:2010,Couch:2014,Couch:2015_1}.

The long-time evolution is important to estimate  observables such as $E_\mathrm{exp}$, $M_\mathrm{ej}$, and $M_\mathrm{PNS}$. We have found that simultaneous accretion and ejection of matter can persist for several seconds after shock revival and delay the saturation of these observables. In particular, the late-time mass accretion onto the PNS is  correlated to the grow of the explosion energy, and governs the directionality of explosion-energy generation. Besides different progenitors, neutrino heating, and rotation strengths, we find that the shock deformation at the onset of the explosion affects the mass accretion of the following seconds. Rapidly rotating models typically explode with a less prolate deformation, suffer less from persistent downflows, and accumulate less explosion energy in the long-time phase. The amount of ejected mass during the long-time phase  is larger in fast rotating  than in  non-rotating models. We note that the prolate/oblate shock deformations in our models are heavily biased by our 2D geometry \citep{Mueller:2015_2,Nakamura:2015,Bruenn:2016,Summa:2016,OConnor:2018_1,Vartanyan:2018}. 

The occurrence of neutrino-driven winds is limited in our models to simulations with the s15 progenitor  and to late times ($t > 3$~s). The total time spent in NDW phases is positively correlated with both the heating factor and rotation rate. However, we also see instances of unstable NDWs and continued accretion after such phases, typically leading to a further increase of the explosion energy. The collapse of a wind is typically induced by renewed accretion from the equatorial plane, which is a fundamentally multi-dimensional effect. Long-time simulations in 3D  are required to further investigate the NDW.

For an analysis of the nucleosynthesis in our models, we used a tracer particle scheme that we applied to all exploding simulations of the 15~$M_\odot$ progenitor. Due to our simplified neutrino treatment, we needed to correct the neutrino energies and luminosities to account for electron fractions that are consistent with modern transport schemes. We have presented the nucleosynthesis for cold trajectories that are not exposed to neutrinos as well as for hot trajectories assuming different electron fraction distributions. All trajectories produce predominately iron group elements with hot, neutrino-processed trajectories reaching elements beyond iron depending on the electron fraction. Our results are a good indication for the potential variability in the production of elements around Sr, Y, Zr and indicate that $^{56}$Ni and $^{44}$Ti are mainly produced by the hot component. In the future, longer simulations times  in three dimensions with detailed neutrino transport will be necessary to connect the nucleosynthesis with the astrophysical conditions (e.g., rotation, explosion energy, progenitor star).

\acknowledgements{
We thank M. Eichler, B. M\"uller, E. O'Connor, and A. Perego for helpful discussions. This research was supported by the ERC Starting Grant EUROPIUM-677912 and Deutsche Forschungsgemeinschaft through SFB 1245. This work has benefited from the COST Action “ChETEC” (CA16117) supported by COST (European Cooperation in Science and Technology) and from JINA Center for the Evolution of the Elements (National Science Foundation under Grant No. PHY-1430152). MO and MR acknowledge the support by the Spanish Ministry of Science, Education and Universities (PGC2018-095984-B-I00) and the Valencian Community (PROMETEU/2019/071). MO further acknowledges support from the Ministerio de Ciencia e Innovación via the Ramón y Cajal program (RYC2018-024938-I). S.M.C. is supported by the U.S. Department of Energy, Office of Science, Office  of Nuclear Physics, Early Career Research Program under Award Number DE-SC0015904. This material is based upon work supported by the U.S. Department of Energy, Office of Science, Office of Advanced Scientific Computing Research and Office of Nuclear Physics, Scientific Discovery through Advanced Computing (SciDAC) program under Award Number DE- SC0017955. This research was supported by the Exascale Computing Project (17-SC-20-SC), a collaborative effort of the U.S. Department of Energy Office of Science and the National Nuclear Security Administration. Calculations for this study were conducted on the Lichtenberg high performance computer of the TU Darmstadt (Project IDs 996 \& 1043).}

\bibliography{main.bib}

\end{document}